\documentclass{article}

\usepackage{nicematrix}
\usepackage{arxiv}

\usepackage[utf8]{inputenc} % allow utf-8 input
\usepackage[T1]{fontenc}    % use 8-bit T1 fonts
\usepackage{hyperref}       % hyperlinks
\usepackage{url}            % simple URL typesetting
\usepackage{booktabs}       % professional-quality tables
\usepackage{amsfonts}       % blackboard math symbols
\usepackage{nicefrac}       % compact symbols for 1/2, etc.
\usepackage{microtype}      % microtypography
\usepackage{lipsum}
\usepackage{graphicx}
\graphicspath{ {./images/} }

\usepackage{mathtools}
\usepackage{multirow}
\usepackage{blindtext}
\usepackage{graphicx}
\usepackage{caption}
\usepackage{physics}
\usepackage{subcaption}
\usepackage{amssymb}
\usepackage{bbm}
\usepackage{amsmath}
\allowdisplaybreaks[1]
\usepackage{tikz}
\usetikzlibrary{shapes,positioning}
\usetikzlibrary{arrows.meta}
\usepackage{float}
\usepackage[utf8]{inputenc}
\usepackage{fancyhdr}
\usepackage{titlesec}
\usepackage{etoolbox}
\usepackage[titletoc]{appendix}
\usepackage{makeidx}
\usepackage{hyperref}
\usepackage{wrapfig}
\makeindex
\usepackage[nottoc]{tocbibind}

\newcommand\numberthis{\addtocounter{equation}{1}\tag{\theequation}}

\patchcmd{\chapter}{\thispagestyle{plain}}{\thispagestyle{fancy}}{}{}

\titleformat{\chapter}{\normalfont\huge\bf}{\thechapter.}{20pt}{\huge}

\newcommand{\beginsupplement}{%
        \setcounter{table}{0}
        \renewcommand{\thetable}{S\arabic{table}}%
        \setcounter{figure}{0}
        \renewcommand{\thefigure}{S\arabic{figure}}%
     }

\title{Misspecification of the generation time  distribution and its impact on $R_t$ estimates in  structured populations}

\author{
 Ioana Bouros \\
  Department of Computer Science\\
  University of Oxford\\
  Oxford, UK \\
  Department of Pathobiology \& Population Sciences\\
  Royal Veterinary College\\ London, UK\\
  \texttt{ibouros@rvc.ac.uk} \\
   \And
 Robin N. Thompson \\
  Mathematical Institute\\
  University of Oxford\\
  Oxford, UK \\
  \texttt{robin.thompson@maths.ox.ac.uk } \\
  \And
 David Gavaghan \\
  Department of Computer Science\\
  University of Oxford\\
  Oxford, UK \\
  \texttt{david.gavaghan@dtc.ox.ac.uk } \\
  \And
  Ben Lambert \\
  Department of Statistics \& Pandemic Sciences Institute\\
  University of Oxford\\
  Oxford, UK \\
  \texttt{ben.lambert@stats.ox.ac.uk} \\
}

\begin{document}
\maketitle
\begin{abstract}
Due to its ability to summarise `real-time' epidemic behaviour, the time-dependent reproduction number, $R_t$, is a useful metric for tracking pathogen transmission and quantifying the effects of interventions during infectious disease outbreaks. The predominant models underlying inferred $R_t$ trajectories are renewal equations, their success owing in part to the relatively few assumptions they require. One necessary assumption is the generation time distribution, which summarises the time periods between infections in infector-infectee transmission pairs. This distribution is typically assumed to be the same across all members of a population. In reality, however, it may vary systematically between population groups. In this study, we consider two $R_t$ inference frameworks based on renewal equation models: one for a single, homogeneous group and another accounting for a structured population. We compare the estimates of $R_t$ generated by the two models and investigate, both analytically and through simulations, under which conditions the conclusions drawn from these modelling paradigms differ. We also demonstrate a methodology for selecting the generation time for the one-group model that correctly encapsulates variations between different population groups; this allows us to use a renewal framework for a one-group model to infer $R_t$ when, in fact, the population is structured.  Finally, we use real epidemic data to demonstrate that practical $R_t$ estimates can differ depending on whether the underlying model is the one-group model or the multi-group model. Our results motivate the need for rigorous collection of detailed epidemic data and consideration of differences between population groups to improve the accuracy of $R_t$ estimates that are used to guide public health policy responses.
\end{abstract}

\section{Introduction}

The time-dependent reproduction number, $R_t$, is defined as the number of secondary infections expected to be generated by an infected case occurring at time $t$ over the course of their infectious period. The most commonly inferred version of $R_t$ is the instantaneous reproduction number, which is based on transmission conditions at time $t$ alone, thereby not accounting for future changes in transmission \cite{Fraser2007, Thompson2019, Cori2013}. A value of $R_t$ below one indicates that the epidemic is expected to decline, while a value above one indicates that the epidemic is expected to grow. Estimates of $R_t$ across a range of values of $t$ can then be used to track the `real-time' behaviour of an epidemic. As public health measures change, the value of $R_t$ will vary as a result \cite{Flaxman2020, Brauner2021}. Likewise, variations in environmental conditions, host behaviours or virus biology \cite{Fraser2007, Nishiura2009, Thompson2020} are also reflected in inferred $R_t$ trajectories.

A common framework for estimating $R_t$ involves assuming that cases occur according to a renewal equation model, and fitting that model to disease incidence time series data \cite{Fraser2007, Nishiura2009, Thompson2019, Flaxman2020, Abbott2020}. Renewal equation models involve the assumption that each new case observed at time $t$ may transmit the pathogen to other members of the population, generating on average $R_t$ secondary cases recorded in the following days' incidence counts. The probability that one of these secondary cases occurs in the disease incidence time series $s$ days after their infector is given by the $s^\text{th}$ entry of the generation time vector $w$ \cite{Svensson2007}. We note that there is some subtlety regarding whether the relevant probability distribution is the generation time distribution (governing the period between infection times in infector-infectee pairs \cite{hart2022inference, hart2022generation}) or the serial interval distribution (governing the period between symptom onset times in infector-infectee pairs \cite{griffin2020rapid}). In reality, neither of these probability distributions exactly reflect the time period between dates of case reporting, particularly when reporting delays are accounted for. However, for simplicity in our terminology, we refer to the `generation time' as the period between cases throughout this article, and verify the robustness of our results for a range of assumed probability distributions.

Renewal equation models have become increasingly popular as a basis for inferring $R_t$ in recent years, likely in part due to the relative simplicity of their setup and the potential for computationally efficient $R_t$ inference. There are many software tools for $R_t$ inference based on renewal equations \cite{Fraser2007, Cori2013, Thompson2020, Abbott2020, Creswell2022}. However, in general, the renewal equations underlying these software tools implicitly assume a population of homogeneous individuals. In alternative modelling frameworks, such as compartmental models (e.g\ \cite{Como-paper, Bouros2024}), individuals in the population are often classified based on factors that are associated with different transmission risks, such as age or vaccination status. Likewise, in theoretical studies, transmission heterogeneity has been included in epidemiological models even in scenarios in which individuals cannot easily be partitioned into different population groups \cite{Lloyd-Smith2005}. Such transmission heterogeneity has been accounted for previously in some studies involving renewal equation-based inference of reproduction numbers (e.g. \cite{johnson2021disease}), but here our focus is on known heterogeneity based on categorisation of individuals into different groups.

A key factor that varies between population groups, in addition to features such as the numbers of contacts between individuals \cite{POLYMOD}, is the generation time. For example, for SARS-CoV-2 in South Korea during the first wave of Omicron variant infections in late 2021, the viral load was estimated to peak two days earlier in children compared to adults \cite{Kim2022}. Since the viral load, in addition to behavioural factors, is linked to transmission, this suggests that generation times may differ depending on the age of the infector. Age typically also has a strong influence on the number of people individuals will interact with on average each day, as well as on the age-distribution of those contacts; in transmission dynamics modelling, contact matrices evaluated from social studies are used to summarise these intricate inter-group dynamics \cite{POLYMOD}.

Some studies in the epidemiological modelling literature have started to recognise and address the fact that the simplest possible renewal equation models routinely used to infer $R_t$ do not account for transmission heterogeneities. However, this work has typically involved additional assumptions about disease dynamics (e.g.\ Glass {\it et al.} \cite{Glass2021} account for the relative differences in the number of contacts across the different population groups) or require additional epidemic quantities to be inferred in order to produce estimates of $R_t$ (e.g\ in \cite{Green2021}, the growth rate of the epidemic is inferred before computing $R_t$ via the next generation matrix approach). 

Previously, in \cite{Bouros2025}, we demonstrated an analytic method for calculating $R_t$ that does not depend on the generation time. Building on the theory derived in \cite{Bouros2025}, we now develop an $R_t$ estimation method that accounts for known structure in the host population. Specifically, we develop a framework for inferring $R_t$ using two distinct models: i) a single group model, in which heterogeneity in transmission between different population groups is not accounted for; ii) a multi-group model, in which transmission heterogeneities are accounted for explicitly. We investigate whether one-group renewal equation models can be suitably adapted for inferring $R_t$ given that, in reality, transmission occurs in heterogeneous populations. We prove analytically that, by considering a generation time distribution for the overall population that averages the group-specific generation times according to the long-term fractions of cases in each population group, estimates of the overall $R_t$ are obtained using the standard one-group renewal equation that match the values of $R_t$ inferred under the multi-group model. Using synthetic epidemic data, we demonstrate that this no longer holds for an epidemic context characterised by continuously changing contact patterns. We also use data from the 2009 outbreak of A/H1N1 in Japan \cite{Nishiura2009b} to show that practical $R_t$ estimates can differ based on whether estimates are inferred under the one-group or multi-group model. It is well-known that the availability of reliable data and frequent data updates are essential to guide outbreak control measures \cite{Thompson2026}. Our results highlight that detailed and representative collection of epidemic data, such as data that can be used to infer group-specific generation time distributions, are required for reliable $R_t$ estimation.

\section{Methods}
\label{multiple-group versus one-group Models}

\subsection{Renewal equation models for structured populations}

In \cite{Bouros2025}, we considered a renewal equation model of pathogen transmission in a population consisting of a single group of individuals. For completeness and ease of reference, we restate the assumptions associated with this modelling approach here. Standard configurations of the renewal equation model assume a stochastic framework to account for the number of cases arising on day $t$, which we denote by $I_t \geq 0$. The number of cases on day $t$ is assumed to depend on the numbers of cases on previous days according to the following model:
\begin{align}
\label{One-group model}
    I_t \sim \text{Poisson}(R_t\Lambda_t)\text{, where } \Lambda_t=\sum^{t-1}_{s=1}w_s I_{t-s},
\end{align}
\noindent where $\Lambda_t$ is the \textit{transmission potential} on day $t$, and $R_t$ is the time-dependent reproduction number at time $t$ for the one-group model \cite{Fraser2007, Thompson2019, Cori2013}. The transmission potential is computed as a weighted average of the number of past cases, where the incidence of cases at time $t-s$ is weighted by $w_s$ -- that is, the $s^\text{th}$ element of the \textit{generation time distribution}. This distribution is discrete with $\sum_{s=1}^\infty w_s=1$ and $w_s \geq 0$ is defined to be the probability that a randomly chosen secondary case occurs $s$ days after the parent case. Also for large $t$, $w_t \approx 0$, so that the vast majority of cases observed are caused by recent parent cases. This model assumes that all infected individuals in the population will mix at equal rates, they are equally likely to transmit the infection and they follow the same timeline of infection progression. This framework, which we refer to as the `one-group' renewal equation model, assumes a homogenously mixing population in which the same generation time distribution is associated with all infected individuals in the population.

However, these assumptions do not necessarily hold in real-world epidemics; for example, children and young adults aged 0-20 interact primarily with other members of their cohort, while older individuals may have on average fewer contacts per day than their younger counterparts, with consequences for epidemiological dynamics \cite{Lovell-Read2022}. We can relax the uniformity assumptions of the standard renewal equation model and integrate some of the mechanisms of population heterogeneity by instead considering a `multi-group' renewal equation approach, similar to the stochastic discrete-time multi-group renewal equation described in \cite{Bouros2025}. In this framework, we model separately the number of cases observed in each population category; the number of new cases on day $t$ in population group $j$ is given by:
\begin{align}
\label{Multiple-group model}
     I_t^{(j)} \sim \text{Poisson}\Big(\gamma_t \sum^{N}_{i=1}C_t^{(ji)}\Lambda_t^{(i)}\Big)\text{, where }\Lambda_t^{(i)}=\sum^{t-1}_{s=1}w_s^{(i)} I_{t-s}^{(i)}.
\end{align}
Cases in group $j$ on day $t$ are generated by previous cases which can belong to any of the $N$ predefined groups in the population. Given a specific case on day $t$, with an infector that was a case $s$ days previously, the probability that the infector belongs to a population group $i$ depends on the $s^\text{th}$ element of the generation time distribution associated with the population group of the infector, that is $w_s^{(i)}$. We use $C_t^{(ji)}$, representing the $(i,j)^\text{th}$ element of a matrix of total effective contacts -- or the average number of effective contacts an individual in group $i$ will have with individuals in group $j$ -- made over the course of an infection following the case at time $t$, to summarise the effects of heterogeneous mixing dynamics of the different population groups. Finally, we assume that all contacts (for an infected individual in a specific group at a specific time in their infection) are equally likely to result in cases, parametrised by the probability a contact becomes a case, $\gamma_t$.

\subsection{Overall and group-specific $R_t$ for multi-group populations}
\label{Rt definitions}

In the following, we identify analytically how the group-specific reproduction numbers of the multiple population group process relate to $R_t$ for the one-group model. We begin by summarising the contribution of previous cases observed in population group $i$ to the number of new cases in group $j$ occurring at time $t$, which we denote as $\mathcal{L}_t^{i \to j}$:
\begin{align}
    \label{eq:group-specific-incidence}
    \mathcal{L}_t^{i \to j} \sim \text{Poisson}\Big(R_t^{i \to j}\Lambda_t^{(i)}\Big).
\end{align}
We use the notation $R_t^{i \to j}$ to represent the average number of secondary cases in group $j$ occurring as a result of one case in group $i$ if conditions remain the same as at time $t$; that is the reproduction number of group $i$ into group $j$ at time $t$. To calculate the reproduction number of the average individual in group $i$, we sum the contributions to subsequent cases across all population categories. The group-specific reproduction number of individuals in group $i$ at time $t$ is then given by $R_t^{(i)} = \sum_{j=1}^N R_t^{i \to j}$. 

The group case incidence in category $j$, $I_t^{(j)}$, is similarly computed by summing the number of new cases in population group $j$ generated by cases in group $i$, $\mathcal{L}_t^{i \to j}$, over all groups. Then, from the properties of Poisson random variables, $I_t^{(j)}$ also follows a Poisson distribution according to:
\begin{align}
\label{Number of New Infection Def Partial 1}
     I_t^{(j)} = \sum^{N}_{i=1}\mathcal{L}_t^{i \to j} \sim \text{Poisson}\Big(\sum^{N}_{i=1}R_t^{i \to j}\Lambda_t^{(i)}\Big).
\end{align}
The total daily disease incidence, $I_t$, is computed by summing $\mathcal{L}_t^{i \to j}$ across all possible population groups $i$ and $j$, that is
\begin{align}
\label{Number of New Infection Definition}
     I_t = \sum^{N}_{i=1}\sum^{N}_{j=1}\mathcal{L}_t^{i \to j} \sim \text{Poisson}\Big(\sum^{N}_{i=1}\sum^{N}_{j=1}R_t^{i \to j}\Lambda_t^{(i)}\Big),
\end{align}
\noindent also from properties of Poisson random variables. By comparing eq.\ \eqref{Number of New Infection Def Partial 1} with eq.\ \eqref{Multiple-group model} governing the group-specific incidence, we arrive at an alternative definition of the reproduction number of group $i$ into group $j$ at time $t$, $R_t^{i \to j}$, which uses instead the total effective number of contacts an infected individual in group $i$ will have in group $j$ over the course of their infection period, $C_t^{(ji)}$, and the probability of a contact becoming a case $\gamma_t$; this between-group reproduction number of group $i$ into group $j$ is given by $R_t^{i \to j} = \gamma_t C_t^{(ji)}$. Hence, eq.\ \ \eqref{Number of New Infection Definition} becomes
\begin{align*}
     I_t &\sim \text{Poisson}\Big(\gamma_t \sum^{N}_{i=1}\sum^{N}_{j=1}{C_t}^{(ji)}\Lambda_t^{(i)}\Big)\\
     &\overset{d}{=} \text{Poisson}\Big( \sum^{N}_{i=1}\gamma_t(\sum^{N}_{j=1}{C_t}^{(ji)})\Lambda_t^{(i)}\Big)\\
     &\overset{d}{=} \text{Poisson}\Big( \sum^{N}_{i=1}R_t^{(i)}\Lambda_t^{(i)}\Big), \numberthis \label{Number of New Infection Def Growth rate}
\end{align*}
\noindent where for the final step we use the identity $R_t^{(i)} = \sum_{j=1}^N R_t^{i \to j} = \sum_{j=1}^N \gamma_t C_t^{(ji)}$ to arrive at a definition of the total case incidence in terms of the group-specific reproduction number $R_t^{(i)}$. Meanwhile, from \cite{Bouros2025}, the overall reproduction number of the multi-group renewal model is defined as $R_t = \gamma_t \rho(C_t)$, where $\rho(\cdot)$ determines the spectral radius, i.e.\ the maximum positive eigenvalue of a matrix, and $C_t$ is the transposed effective contact matrix with entries $\{C_t^{(ij)}\}$. If there are no differences in the transmission between population groups, that is, the group-specific reproduction number $R_t^{(k)}$ and generation times $w_s^{(k)}$ are identical across all groups $k$, we can reduce the multi-group model to the one-group renewal equation model described by eq.\  \eqref{One-group model} if the means of the two processes are equal, i.e.\
\begin{align*}
\gamma_t\sum^{N}_{i=1}\sum^{N}_{j=1}{C_t}^{(ji)}\Lambda_t^{(i)} = R_t\Lambda_t.
\end{align*}
Replacing the probability of a contact becoming a case, $\gamma_t$, by its definition in terms of the group-specific reproduction number in population group $k$, that is $R_t^{(k)} \sum_{j=1}^N \gamma_t C_t^{(jk)}$, the equation above becomes
\begin{align*}
    R_t^{(k)} \frac{1}{\sum_{j=1}^N {C_t}^{(jk)}}\Big(\sum^{N}_{i=1}\sum^{N}_{j=1}{C_t}^{(ji)}\Lambda_t^{(i)}\Big) = R_t\sum^{N}_{i=1}\Lambda_t^{(i)},
\end{align*}
\noindent where $R_t$ is defined as the overall time-dependent reproduction number for the one-group model. The group-specific time-dependent reproduction number $R_t^{(k)}$ in the multi-group renewal equation model can then be defined in terms of the overall time-dependent reproduction number $R_t$ in the one-group population model according to:
\begin{align}
\label{R number perfect contact}
    R_t^{(k)} = R_t \frac{\Big(\sum^{N}_{i=1}\Lambda_t^{(i)}\Big)\Big(\sum_{j=1}^N {C_t}^{(jk)}\Big)}{\sum^{N}_{i=1}\sum^{N}_{j=1}{C_t}^{(ji)}\Lambda_t^{(i)}}.
\end{align}

\subsection{Reproducing our results}

We extend the infrastructure of the `branchpro' software package described in \cite{Creswell2022} to allow forward simulation of renewal equation models for populations with multiple groups, with a user-specified contact matrix as an input.

We also provide `Stan' \cite{Stan} scripts for the inference of the group-specific reproduction number trajectories for both the multi-group and the one-group renewal equation processes using Stan's default No U-Turn sampling algorithm. In order to assess the accuracy of the estimates, we use the $\hat{R}$ statistic (introduced by Gelman and Rubin in \cite{Gelman1992}), and assume that a threshold value of $\hat{R} \lesssim 1.01$ \cite{Vehtari2021} indicates adequate convergence of the MCMC chains. All code required to reproduce the results in this article is available in our public Github repository \cite{branchpro}.

\section{Results}

In the following sections, we derive the posterior distributions for $R_t$ under the one-group and multi-group renewal equation modelling frameworks. We compare the accuracy of our estimates for the overall and group-specific $R_t$ generated using the multi-group renewal equation model with the $R_t$ estimates obtained from the one-group model. We use a range of different simulation scenarios, with a particular focus on how the choice of generation time distributions affects the inferred $R_t$ trajectories. We use Bayesian inference to estimate $R_t$, and we describe our choice of priors in this section.

\subsection{Posterior distribution computation: the one-group and the multi-group models}
\label{Posterior comparison}

We first determine the analytical posterior distribution for the overall reproduction numbers obtained under both the one-group and multi-group renewal equation frameworks. We use the posteriors to investigate under which conditions the overall reproduction numbers identified by the two models are identical. This sheds light on when it is justified to use the simpler one-group approach instead of the multi-group model, which requires more detailed group-specific epidemiological data.

For both the one-group and the multi-group models we consider a $\lambda$-day sliding-window \cite{Thompson2019} when inferring the time-dependent reproduction number. This approach assumes that over any interval of $\lambda$ consecutive days, the reproduction number remains constant, and so information is pooled across this period. For each subsequent $R_t$ value the sliding window is shifted one day to the right, so that the value of $R_t$ is reflective of observed cases on day $t$ and on the $\lambda - 1$ days previously. This way, we avoid sharp fluctuations in the inferred values of $R_t$, which can occur if $R_t$ is estimated based on recipient cases on day $t$ alone (in which case, $R_t$ estimates may simply reflect the noise in the transmission process).

For the single-group renewal equation model, the likelihood function is then given by \cite[eq.~(2), Web Appendix 1]{Cori2013}:
\begin{align*}
    L(R_t| \underline{w}, \underline{I_t}) &= p (I_{t-\lambda + 1}, I_{t-\lambda + 2}, \dots, I_t | \underline{w}, R_t, I_0, \dots, I_{t-\lambda}) =\\
    &= \prod_{k=t-\lambda+1}^t \text{Poisson}\Big(I_k; R_t \sum_{s=1}^{k-1} w_s I_{k-s}\Big),
\end{align*}
\noindent where $\underline{I_t}$ represent the vector of all case incidences until time $t$ -- that is $\underline{I_t} = \{ I_0, I_1, \dots I_t \}$.

For each $R_t$, we assume a gamma distribution as our prior: $p(R_t) = \text{gamma}(R_t; \alpha, \beta)$, where $\alpha$ is the shape and $\beta$ is the scale parameter of the distribution, i.e.\ it has a mean of $\alpha\beta$ and variance of $\alpha\beta^2$. Using Bayes' rule we derive the posterior for $R_t$:
\begin{align*}
    p(R_t| \underline{I_t}) 
    &\propto L(R_t| \underline{I_t}) p(R_t)\\
    &\propto \Bigg(\prod_{k=t-\lambda+1}^t \frac{(R_t \Lambda_k)^{I_k}}{I_k!} e^{-R_t \Lambda_k}\Bigg) \frac{R_t^{\alpha-1}e^{- R_t/\beta}}{\Gamma(\alpha)\beta^\alpha}\\
    &\propto R_t^{\sum_{k=t-\lambda+1}^t I_k} e^{-R_t \sum_{k=t-\lambda+1}^t \Lambda_k}R_t^{\alpha-1}e^{- R_t/\beta}\\
    &\propto R_t^{\alpha - 1 + \sum_{k=t-\lambda+1}^t I_k} \exp{\bigg(- \frac{R_t}{\beta} -R_t \sum_{k=t-\lambda+1}^t \Lambda_k\bigg)}\\
    &\propto R_t^{\alpha - 1 + \sum_{k=t-\lambda+1}^t I_k} \exp{\bigg(- R_t\Big(\frac{1}{\beta} + \sum_{k=t-\lambda+1}^t \Lambda_k\Big)\bigg)},\\
    \text{i.e.\ }R_t| \underline{I_t}&\sim \text{gamma}\Bigg(\alpha + \sum_{k=t-\lambda+1}^t I_k, \Big(\frac{1}{\beta} + \sum_{k=t-\lambda+1}^t \Lambda_k\Big)^{-1}\Bigg).
    \numberthis \label{one-group Posterior}
\end{align*}

We now determine the posterior distribution of the overall reproduction number $R_t$ for a multi-group renewal model. As described above, the overall reproduction number is given by $R_t= \gamma_t \rho(C_t)$. Similarly to the one-group renewal equation modelling framework, the value of the overall reproduction number determines the long-term behaviour of the trajectory of new cases. We rewrite the renewal equation governing the number of cases observed in population group $j$ on day $t$ from eq.\  \eqref{Multiple-group model} in terms of the overall reproduction number:
\begin{align*}
    I_t^{(j)} \sim \text{Poisson}\Big(\frac{R_t}{\rho(C_t)} \sum^{N}_{i=1}{C_t}^{(ji)}\Lambda_t^{(i)}\Big).
\end{align*}

\noindent We then aggregate the number of new cases recorded on day $t$ in each of the $N$ population categories and arrive at an equivalent definition of the multi-group renewal equation model:
\begin{align*}
    I_t \sim \text{Poisson}\Big(\frac{R_t}{\rho(C_t)} \sum^{N}_{j=1}\sum^{N}_{i=1}{C_t}^{(ji)}\Lambda_t^{(i)}\Big).
\end{align*}

We use the same prior distribution for $R_t$ as in the one-group case. Applying Bayes' rule once again, the posterior distribution for the multi-group model is also a gamma-distribution with the following form:
\begin{align*}
    p(R_t| \underline{I_t}) 
    \propto& L(R_t| \underline{I_t}) p(R_t)\\
    \propto& R_t^{\alpha - 1 + \sum_{k=t-\lambda+1}^t I_k} \exp{\bigg(- R_t\Big(\frac{1}{\beta} + \frac{1}{\rho(C_t)} \sum_{k=t-\lambda+1}^t \sum^{N}_{j=1}\sum^{N}_{i=1}{C_k}^{(ji)}\Lambda_t^{(i)}\Big)\bigg)},\\
    \text{i.e., }R_t| \underline{I_t} \sim& \text{gamma}\Bigg(\alpha + \sum_{k=t-\lambda+1}^t I_k, \Big(\frac{1}{\beta} + \frac{1}{\rho(C_t)} \sum_{k=t-\lambda+1}^t \sum^{N}_{j=1}\sum^{N}_{i=1}{C_k}^{(ji)}\Lambda_t^{(i)}\Big)^{-1}\Bigg).
    \numberthis \label{multiple-group Posterior}
\end{align*}

\subsection{Differences in the overall $R_t$ estimates}
In this section, we investigate how selecting the generation time distribution for the overall population when assuming a one-group population renewal equation changes the inferred overall $R_t$ profile. A naive candidate for the one-group generation time  would be to compute the average over the group-specific generation times distribution. As we demonstrate below, this choice does not always lead to accurate overall $R_t$ estimates.

\begin{figure}[ht!]
  % include first image
  \includegraphics[width=1\linewidth]{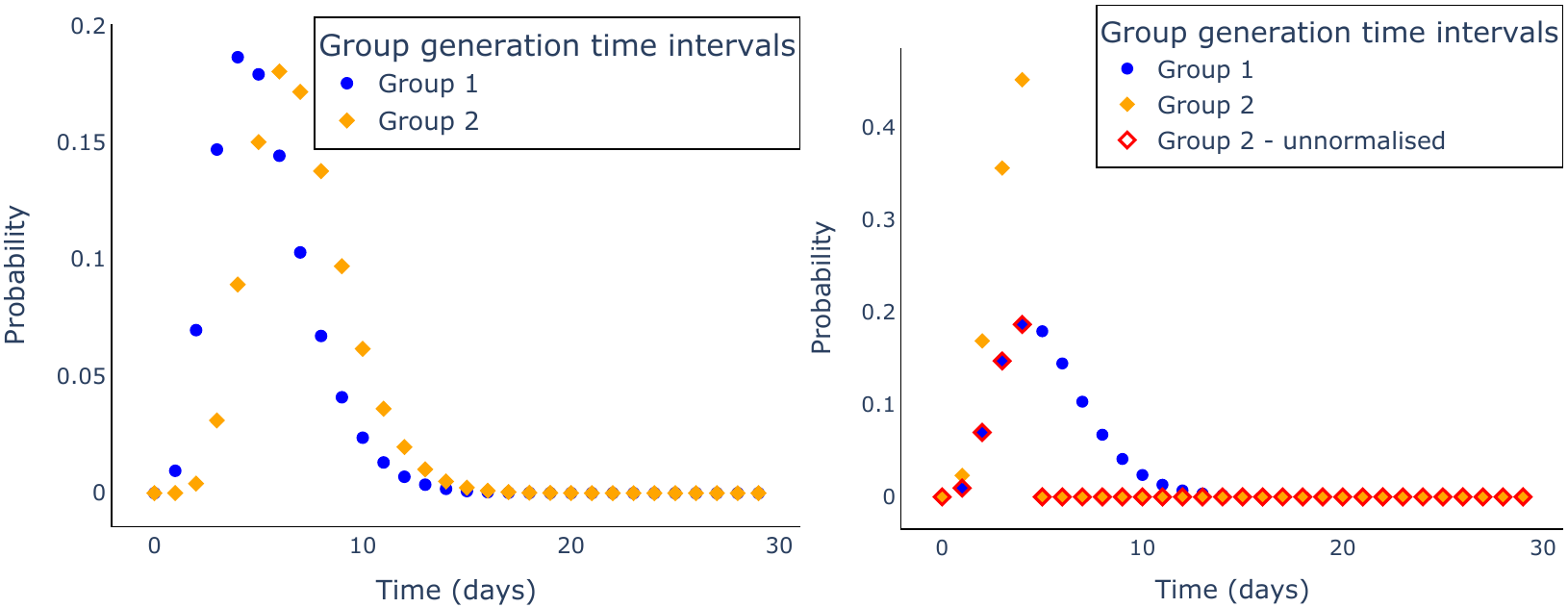}
\caption{\textbf{Group-specific generation time distributions used in our analyses.} Generation time distributions are shown for a two-group toy population when: (left-panel) they match up to day $5$ in their unnormalised form, (right-panel) they are different throughout. These generation times follow a discretised gamma distribution with mean and standard deviation as described in Table \ref{tab:generation-time-comparison}.}
\label{Group-specific generation times}
\end{figure}

We begin by assuming that we are modelling a population with two groups. For each of these groups, the generation time distributions can either (i) be identical, (ii) partially match (the left panel of Figure \ref{Group-specific generation times}) -- this could reflect, for example, the behaviour of a risk-averse group that chooses to self-isolate when symptoms appear -- or (iii) differ persistently between the two population groups (right panel of Figure \ref{Group-specific generation times}). These generation time distributions are presented in their discretised versions, computed using the method presented in \cite{Cori2013, ogi2025real, ogi2025simulation} and are normalised to sum to one. We investigate how the mean inferred overall $R_t$ trajectory differs between the one-group and the multi-group models respectively for each of these three generation time distribution scenarios. We use the properties of the analytical form of the posterior distributions computed in eqs.\ \eqref{one-group Posterior} \& \eqref{multiple-group Posterior} to produce the estimated trajectory (and their $95\%$ confidence regions) of the overall $R_t$ for each of the two models studied: the one-group model and the multi-group renewal model, respectively.

In Tables \ref{tab:generation-time-comparison} and \ref{tab:comparison-conditions}, we summarise the chosen parameters for each of the scenarios considered. The choice of parameters is for illustrative purposes only and are not indicative of any particular population or disease -- we are more interested in visualising how these changes qualitatively impact the estimates of $R_t$, rather than modelling any particular epidemic.
\begin{table}[ht!]
    \centering
\renewcommand{\arraystretch}{1.3}
\begin{tabular}{|>{\centering\arraybackslash}m{2cm}|>{\centering\arraybackslash}m{3cm}|>{\centering\arraybackslash}m{3cm}|>{\centering\arraybackslash}p{3cm}|>{\centering\arraybackslash}p{3cm}|}
\hline
         & \multicolumn{2}{|c|}{\textbf{Mean generation time}}&   \multicolumn{2}{|c|}{\textbf{Standard deviation of generation time}}\\
\hline
\textbf{Scenario} & Group 1& Group 2& Group 1& Group 2\\
\hline
 Different& 5.3& 7& 2.3& 2.3\\ \hline 
Same, cropped& 5.3& 5.3 (before cropping) 3.24 (after cropping)& 2.3& 2.3 (before cropping) 0.81 (after cropping)\\\hline 
Same& 5.3& 5.3& 2.3& 2.3\\ \hline
\end{tabular}
\vspace{3mm}
\caption{Generation time distributions used in our simulation scenarios. All values  indicate times measured in days.}\label{tab:generation-time-comparison}
\end{table}
\begin{table}[ht!]
    \centering
\renewcommand{\arraystretch}{1.3}
\begin{tabular}{|c|c|c|}
\hline 
\textbf{Parameter}&  \textbf{Original}& \textbf{Changed}\\ \hline 
         Contact matrix ($C_t$)&  $\begin{pmatrix} 6 & 8\\ 
4 & 8 \end{pmatrix}$ & $\begin{pmatrix}
0.4 & 4\\ 
6 & 8
\end{pmatrix}$ \\ \hline 
         Probability of a contact becoming a case ($\gamma_t$)&  [0.35, 0.05, 0.4, 0.05]& [0.35, 0.05, 0.2, 0.1]\\ \hline
 Times of changes in $\gamma_t$& [0, 37, 60, 160]&[0, 37, 60, 160]\\\hline  
         Initial Conditions&  [50, 50]& [20, 80]\\ \hline
    \end{tabular}
    \vspace{3mm}
    \caption{Model parameters used across in simulation scenarios.}
    \label{tab:comparison-conditions}
\end{table}

To determine whether our results hold generally, we repeat our analyses, changing each of the other model parameters one at a time and plot the results in Figures \ref{Same serial interval}, \ref{Same cropped serial interval} and \ref{Different serial interval}; in each of the Figures \ref{Same serial interval}, \ref{Same cropped serial interval} and \ref{Different serial interval} we consider changes in the contact matrix $C_t$ (top right panel), the probability of a contact becoming a case $\gamma_t$ (bottom left panel) or initial number of cases in each population group (bottom right panel). In Table \ref{tab:comparison-conditions}, we record the chosen values of these other model parameters. The underlying cases datasets we fitted the models to in each parameterisation regime considered were generated using a two-group renewal equation model, as described in eq.\ \eqref{Multiple-group model}, and are plotted in Figures \ref{Same SI Incidence}, \ref{Same cropped SI Incidence} and \ref{Different SI Incidence}.

When all population groups share the same generation times, the $R_t$ trajectory inferred using the one-group renewal equation modelling framework matches that of the overall average $R_t$ from the multi-group model, at least after an initial time delay corresponding to the system responding to the initial conditions (Figure \ref{Same serial interval}). However, this is no longer the case if there are any differences in the generation time distributions between the two population groups (Figure \ref{Same cropped serial interval} and Figure \ref{Different serial interval}). Moreover, these differences in the estimates of $R_t$ do not follow a consistent pattern; using the one-group model instead of the multi-group model can both lead to overestimates (Figure \ref{Different serial interval}, top left panel) and underestimates (Figure \ref{Different serial interval}, top right panel) of $R_t$, depending on the contact matrix and other model parameters.

Having illustrated this misalignment of $R_t$ estimates through simulations, we now demonstrate analytically the generality of the result: for any population with $N$ groups with the same generation time, the overall inferred $R_t$ trajectory perfectly matches the one inferred using the one-group population model with the same generation time in the long run. The discordance observed at the beginning of the dataset does not persist as the simulation continues. 

First, by comparing the forms of the posteriors in eq.\ \eqref{one-group Posterior} and eq.\ \eqref{multiple-group Posterior}, we observe that both share the same shape parameter. Hence, the overall reproduction number trajectories produced using the one-group and multi-group models are identical when the scale parameters of the gamma distributions match. This holds if and only if
\begin{align*}
\allowdisplaybreaks
\begin{aligned}
    \sum_{k=t-\lambda+1}^t \sum_{s=1}^{k-1} w_s I_{k-s} &= \frac{1}{\rho(C_t)} \sum_{k=t-\lambda+1}^t \sum^{N}_{j=1}\sum^{N}_{i=1}{C_k}^{(ji)}\sum^{k-1}_{s=1}w_s^{(i)} I_{k-s}^{(i)}.
\end{aligned}
\end{align*}
We consider the ratio between the means of the two posteriors as a measure of the divergence of the (more realistic) multi-group renewal equation model from the simplified one-group model. Since the mean of a $\text{gamma}(\alpha, \beta)$ distribution is given by the product of the shape parameter $\alpha$ and scale parameter $\beta$, and the shape parameter is identical between the two models, the ratio evaluated at time $t$ becomes equal to:
\begin{align*}
    \text{Ratio}(t) = \frac{\begin{gathered}
    \text{multiple-group} \\ \text{mean}
    \end{gathered}}{
    \begin{gathered}
    \text{one-group} \\ \text{mean}
    \end{gathered}}
    &= \frac{\frac{1}{\beta} + \sum_{k=t-\lambda+1}^t \sum_{s=1}^{k-1} w_s I_{k-s}}{\frac{1}{\beta} + \frac{1}{\rho(C_t)} \sum_{k=t-\lambda+1}^t \sum^{N}_{j=1}\sum^{N}_{i=1}{C_k}^{(ji)}\sum^{k-1}_{s=1}w_s^{(i)} I_{k-s}^{(i)}}.
    \numberthis \label{Ratio}
\end{align*}
\noindent Identical generation times in all population groups imply that $w_s^{(i)}=w_s, \forall i$, for all times $s$. If mixing within and between groups is identical and the groups are of the same size, there are equal numbers of daily contacts in all groups ($C_t=\frac{m}{N}\mathbb{J}_N$, where each individual is expected to have $m$ effective contacts per day in total, and where $\mathbb{J}_N$ denotes the $N \times N$ matrix with all elements equal to one). Thus, the ratio defined in eq.\ \eqref{Ratio} is exactly one for all times $t$, because:
\begin{align*}
    \frac{1}{\rho(\frac{m}{N}\mathbb{J}_N)} \sum_{k=t-\lambda+1}^t \sum^{N}_{j=1}\sum^{N}_{i=1}\frac{m}{N}\sum^{k-1}_{s=1}w_s I_{k-s}^{(i)} &= \frac{1}{m}\sum_{k=t-\lambda+1}^t \sum^{N}_{i=1}\bigg(\sum^{N}_{j=1}\frac{m}{N}\bigg)\sum^{k-1}_{s=1}w_s I_{k-s}^{(i)}=\\
    = \frac{1}{m}\sum_{k=t-\lambda+1}^t \sum^{N}_{i=1} m \sum^{k-1}_{s=1}w_s I_{k-s}^{(i)}
    &=\frac{m}{m} \sum_{k=t-\lambda+1}^t \sum^{k-1}_{s=1}w_s \bigg(\sum^{N}_{i=1} I_{k-s}^{(i)}\bigg) = \sum_{k=t-\lambda+1}^t \sum^{k-1}_{s=1}w_s I_{k-s}.
\end{align*}
However, for a more general form of the transposed effective contact matrix, we can show that the ratio in eq.\ \eqref{Ratio} approaches one for large enough time $t$. At this point, the epidemic will reach a state where the fractions of cases in each group either grow or decay exponentially with the same growth rate across all groups. This is the long-term solution and, for the discrete time renewal equation process, we can write this as follows (from \cite[appendix~ A.6]{Bouros2025}):
\begin{align*}
    \begin{pmatrix}
I^1_t \\
\dots \\
I^N_t 
\end{pmatrix} = (1+r)^t \underline{\Phi},
\end{align*}
\noindent where $r$ is the exponential outbreak growth rate and $\underline{\Phi}$ is the vector of the number of new cases in each population group once the systems reaches stability in terms of the fractions of cases in each group. According to \cite{Bouros2025}, the equilibrium vector $\underline{\Phi}$ satisfies:
\begin{align}
\label{Laplace Transform}
    \underline{\Phi} &= \Big(\sum_{s=0}^\infty \gamma_t C_t W(s) (1+r)^{-s} \Big) \underline{\Phi} = \gamma_t C_t \Big(\sum_{s=0}^\infty W(s) (1+r)^{-s} \Big) \underline{\Phi} := \overline{K}(r) \underline{\Phi},
\end{align}
\noindent where $\overline{K}(r)$ is the Laplace Integral transform of $\gamma_t C_t W(s)$ and $W(s)$ is the diagonal matrix of the generation times evaluated at time $s$, for all population groups -- that is for any continuous time point $s$, we have
$$W(s) = \begin{bmatrix}
    w^1(s) & \dots & 0\\
    \vdots & \ddots & \vdots\\
    0 & \dots & w^N(s)
  \end{bmatrix},$$
\noindent where $w^i(s)$ is the evaluation at the time $s$ of the continuous time version of the discrete generation interval of group $i$, $w^i_s$. We have also shown in \cite{Bouros2025} that the spectral radius of the Laplace Integral transform of $\gamma_t C_t W_s$ satisfies $\rho(\overline{K}(r))=1$. Since all population groups share the same generation time , $W(s)$ satisfies $W(s) = w_s \mathbb{I}_N$ for all time points $s$, where $\mathbb{I}_N$ denotes the $N \times N$ identity matrix. Applying this property to eq.\ \eqref{Laplace Transform}, we get that:
\begin{align*}
    \underline{\Phi} &= \Big(\sum_{s=0}^\infty \gamma_t C_t W(s) (1+r)^{-s} \Big) \underline{\Phi} = \Big(\sum_{s=0}^\infty \gamma_t C_t w_s \mathbb{I}_N (1+r)^{-s} \Big) \underline{\Phi} = \gamma_t \Big(\sum_{s=0}^\infty w_s (1+r)^{-s} \Big) C_t \underline{\Phi}, \numberthis \label{Phi}
\end{align*}
\noindent and the spectral radius of the transposed effective contact matrix $C_t$ then satisfies:
\begin{align*}
    1 &= \rho(\overline{K}(\gamma)) = \rho\Bigg(\gamma_t \Big(\sum_{s=0}^\infty w_s (1+r)^{-s} \Big) C_t \Bigg) = \gamma_t \Big(\sum_{s=0}^\infty w_s (1+r)^{-s} \Big) \rho(C_t). \numberthis \label{rho Phi}
\end{align*}
\noindent Substituting the results from equations \eqref{Phi} and \eqref{rho Phi} into the ratio coefficient defined in eq.\ \eqref{Ratio}, we conclude that for large times $t$ we have:
\begin{align*}
    \text{Ratio}(t) &=
    \frac{\frac{1}{\beta} + \sum_{k=t-\lambda+1}^t \sum_{s=1}^{k-1} w_s \sum^{N}_{j=1} I_{k-s}^{(j)}}{\frac{1}{\beta} + \frac{1}{\rho(C_t)} \sum_{k=t-\lambda+1}^t \sum^{N}_{j=1}\sum^{N}_{i=1}{C_t}^{(ji)}\sum^{k-1}_{s=1}w_s I_{k-s}^{(i)}}\\
    &= \frac{\frac{1}{\beta} + \sum_{k=t-\lambda+1}^t \sum_{s=1}^{k-1} w_s \sum^{N}_{j=1} (1+r)^{(k-s)}\phi^{(j)}}{\frac{1}{\beta} + \frac{1}{\rho(C_t)} \sum_{k=t-\lambda+1}^t \sum^{N}_{j=1}\sum^{N}_{i=1}{C_t}^{(ji)}\sum^{k-1}_{s=1}w_s (1+r)^{(k-s)}\phi^{(i)}}\\
    &= \frac{\frac{1}{\beta} + (\sum_{k=t-\lambda+1}^t \sum_{s=1}^{k-1} w_s (1+r)^{(k-s)}) (\sum^{N}_{j=1} \phi^{(j)})}{\frac{1}{\beta} + (\sum_{k=t-\lambda+1}^t \sum^{k-1}_{s=1}w_s (1+r)^{(k-s)})(\frac{1}{\rho(C_t)}\sum^{N}_{j=1}\sum^{N}_{i=1}{C_t}^{(ji)}\phi^{(i)})}. \numberthis \label{Ratio with Phis}
\end{align*}

Summing across all population groups in eq.\ \eqref{Phi}, we observe that the total number of cases when the system reaches the equilibrium state vector $\underline{\Phi}$ satisfies:
\begin{align*}
\sum^{N}_{j=1} \phi^{(j)} = \gamma_t \Big(\sum_{s=0}^\infty w_s (1+r)^{-s} \Big) \sum^{N}_{j=1} \sum^{N}_{i=1} {C_t}^{(ji)} \phi^{(i)} = \frac{1}{  \rho(C_t)} \sum^{N}_{j=1} \sum^{N}_{i=1} {C_t}^{(ji)} \phi^{(i)},
\end{align*}
\noindent which when substituted into eq. \eqref{Ratio with Phis} simplifies the fraction to exactly one. Therefore, when all population groups have the same generation times, the one-group and multi-group $R_t$ estimates are equivalent for large time $t$.

We now explore the circumstances when the generation time distribution differs across groups. We investigate if it is possible to still use the one-group model to infer the correct $R_t$ trajectory by using a weighted generation time distribution.

From eq.\ \eqref{Ratio}, we have that the posterior distribution for $R_t$ at time $t$ for the one-group and multi-group models match if
\begin{align*}
\sum_{k=t-\lambda+1}^t \sum_{s=1}^{k-1} w_s I_{k-s} &= \sum_{k=t-\lambda+1}^t \sum^{k-1}_{s=1} \frac{1}{\rho(C_t)} \sum^{N}_{j=1}\sum^{N}_{i=1}{C_k}^{(ji)}w_s^{(i)} I_{k-s}^{(i)}.
\numberthis \label{Ratio Different ws}
\end{align*}
As seen from eq. \eqref{Phi}, the long-run incidence of cases in population group $i$ is given by $I^{(i)}_{k-s} = (1+r)^{(k-s)}\phi^{(i)}$, where $\phi^{(i)}$ satisfies
\begin{align}
\underline{\Phi} &= \Big(\sum_{s=0}^\infty \gamma_t C_t W(s) (1+r)^{-s} \Big) \underline{\Phi}.
\label{Phi eq}
\end{align}
\noindent Substituting eq.\ \eqref{Phi eq} into eq.\ \eqref{Ratio Different ws} removes the dependency on $k$, if the number of contacts (that is the contact matrix $C_k$) remains constant over time. Hence, a generation times distribution for the one-group model that satisfies
\begin{align}
w_s= \frac{1}{\rho(C_t)} \frac{\sum^{N}_{j=1}\sum^{N}_{i=1}{C_t}^{(ji)}w_s^{(i)} I_{k-s}^{(i)}}{I_{k-s}} = \sum^{N}_{i=1} \frac{\sum^{N}_{j=1}{C_t}^{(ji)}}{\rho(C_t)}w_s^{(i)} \frac{\phi^{(i)}}{\sum^{N}_{l=1} \phi^{(l)}},
\label{one-group correct generation time}
\end{align}
\noindent will ensure equality between the posteriors for $R_t$ under the one-group and the multi-group renewal equation models. The fraction $\frac{\phi^{(i)}}{\sum^{N}_{l=1} \phi^{(l)}}$ represents the long-term proportion of cases in population group $i$ and is determined by the eigenvector corresponding to the maximum eigenvalue of the matrix $M=\gamma_t C_t \sum_{a=0}^\infty W(a) (1+r)^{a}$, which is one.

For a two-group population, we demonstrate an exact formula for the overall generation time distribution. Based on the above, we have that
\begin{align*}
\begin{pmatrix}
M^{(11)} & M^{(12)} \\
M^{(21)} & M^{(22)}
\end{pmatrix}
\begin{pmatrix}
\phi^{(1)} \\
\phi^{(2)}
\end{pmatrix} = \begin{pmatrix}
\phi^{(1)} \\
\phi^{(2)}
\end{pmatrix}.
\end{align*}
\noindent This equality can be reinterpreted in terms of the long-term proportion of cases in each of the population groups $1$ and $2$ as
\begin{align*}
\phi^{(1)} &= M^{(11)}\phi^{(1)} + M^{(12)}\phi^{(2)} = C_t^{(11)}\phi^{(1)} \sum_{a=0}^\infty w^{(1)}(a) (1+r)^{-a} +  C_t^{(12)}\phi^{(2)} \sum_{a=0}^\infty w^{(2)}(a) (1+r)^{-a},\\
\phi^{(2)} &= M^{(21)}\phi^{(1)} + M^{(22)}\phi^{(2)}= C_t^{(21)}\phi^{(1)} \sum_{a=0}^\infty w^{(1)}(a) (1+r)^{-a} +  C_t^{(22)}\phi^{(2)} \sum_{a=0}^\infty w^{(2)}(a) (1+r)^{-a}.
\end{align*}
\noindent By summing these, we obtain
\begin{align*}
\phi^{(1)}+\phi^{(2)} = (C_t^{(11)} + C_t^{(21)})\phi^{(1)} \sum_{a=0}^\infty w^{(1)}(a) (1+r)^{-a} + (C_t^{(12)} + C_t^{(22)})\phi^{(2)} \sum_{a=0}^\infty w^{(2)}(a) (1+r)^{-a}.
\end{align*}
The correctly weighted one-group generation time  from eq.\ \eqref{one-group correct generation time} then becomes
\begin{align}
\label{Corrected GT}
    w_s = \frac{(C_t^{(11)}+C_t^{(21)}) \phi^{(1)} w_s^{(1)}+ (C_t^{(12)}+C_t^{(22)}) \phi^{(2)} w_s^{(2)}}{\rho(C_t)(\phi^{(1)} + \phi^{(2)})}.
\end{align}
\subsection{Conditions under which the $R_t$ estimates from the two-group population model align with those of a one-group model}

We now explore how our theoretical result (that using a carefully chosen generation time distribution in the one-group model facilitates $R_t$ inference that matches estimates from the multi-group model) applies when $R_t$ is inferred from disease incidence time series data. To do this, we consider the same simulation conditions as before (see Tables \ref{tab:generation-time-comparison} and \ref{tab:comparison-conditions} for the model parameters in each population scenario), with the slight amendment that the  probability of a contact becoming a case $\gamma_t$ only changes once at $t=37$, as shown in Table \ref{tab:correct_comparison-conditions}.
\begin{table}[H]
    \centering
\renewcommand{\arraystretch}{1.3}
\begin{tabular}{|c|c|c|}
\hline 
\textbf{Parameter}&  \textbf{Original}& \textbf{Changed}\\ \hline 
         Probability of a contact becoming a case ($\gamma_t$)&  [0.35, 0.2]& [0.35, 0.38]\\ \hline
 Times of changes in $\gamma_t$ & [0, 37] & [0, 37]\\\hline
    \end{tabular}
    \vspace{3mm} \caption{Probability of a contact becoming a case used in our simulation scenarios.}\label{tab:correct_comparison-conditions}
\end{table}

\begin{figure}[ht]
  \centering
  % include first image
  \includegraphics[width=.7\linewidth]{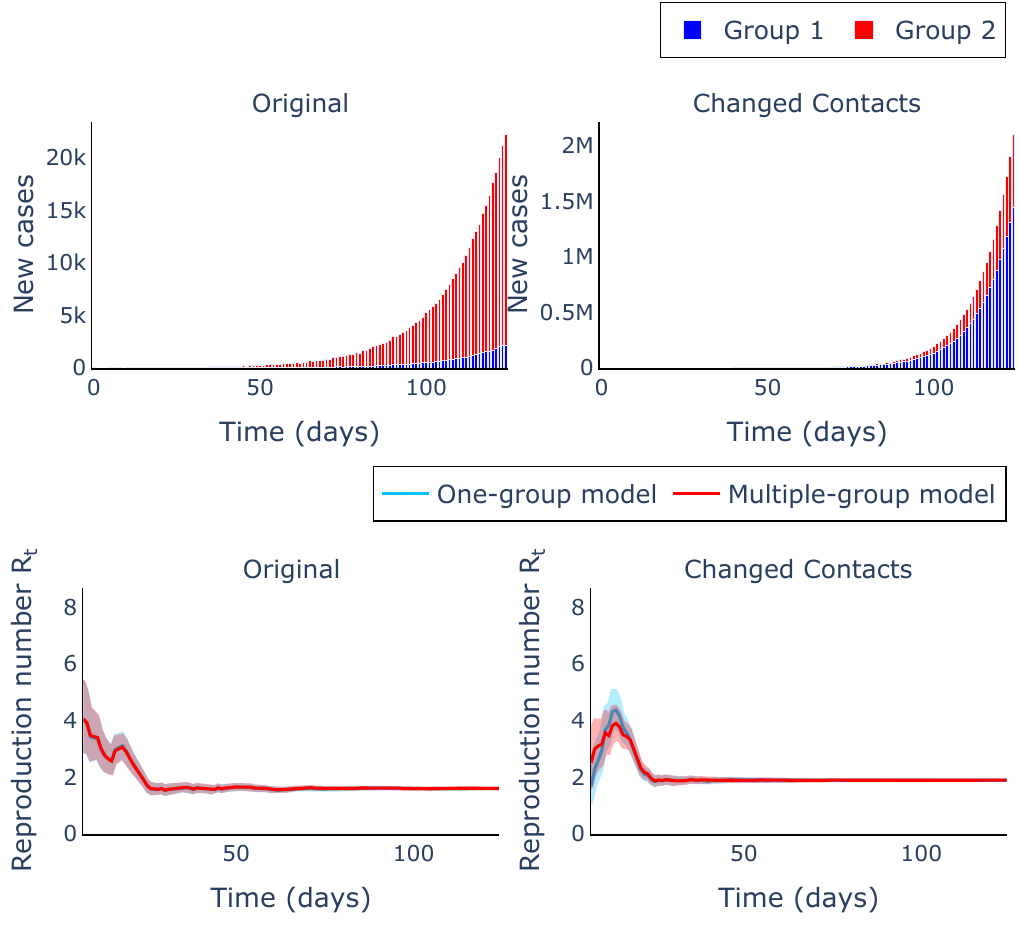}
  \vspace{0.05mm}
\caption{\textbf{$R_t$ estimates obtained from the one- and multi-group models match in the long-term when the contact matrix does not vary temporally.} The mean and 95\% confidence region of the overall $R_t$ trajectories are inferred using the multi-group model (red lines) and one-group model (light blue lines) for a range of model parameters as described in Table \ref{tab:comparison-conditions}, when the generation time for the one-group model is calculated appropriately (as described in the text). The simulated disease incidence data (top panels) underlying the $R_t$ estimates are generated using a two-group model.}
\label{Corrected serial interval}
\end{figure}
\begin{figure}[ht!]
  \centering
  % include first image
  \includegraphics[width=.7\linewidth]{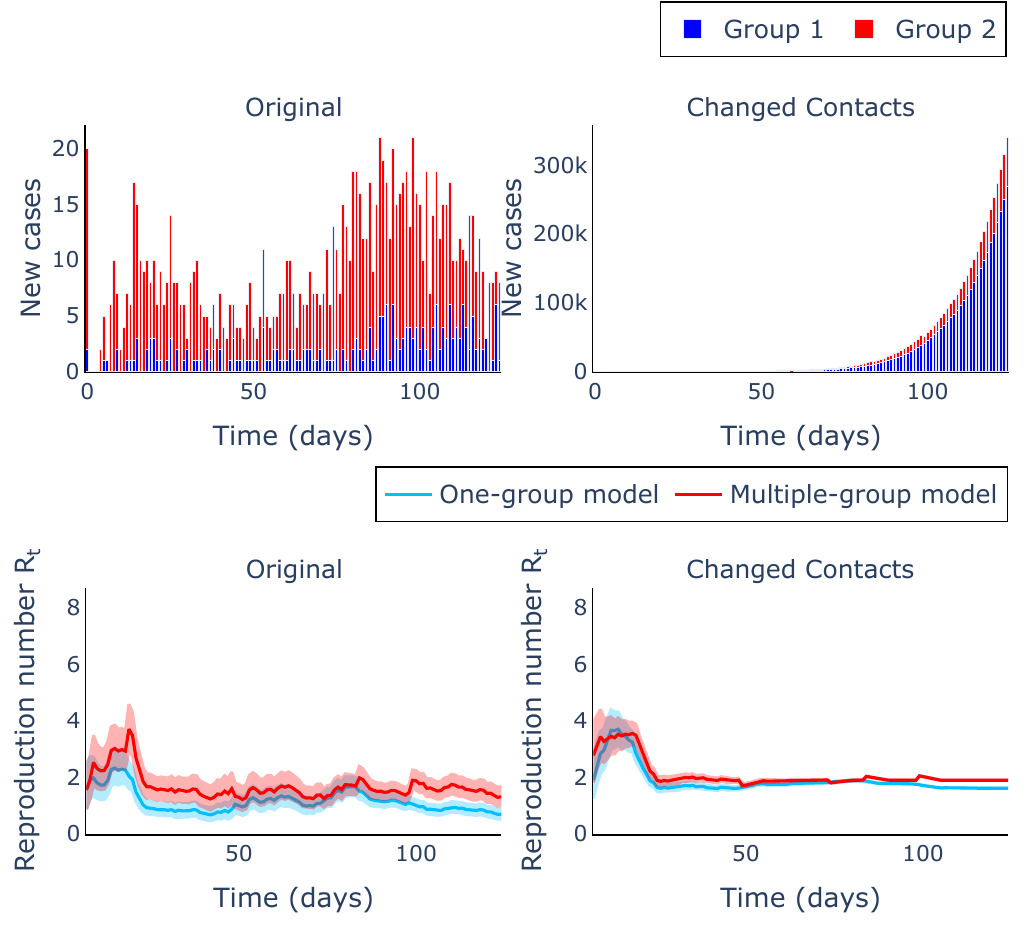}
  \vspace{0.05mm}
\caption{\textbf{$R_t$ estimates obtained from the one- and multi-group models diverge when the contact matrix varies temporally.} The mean and 95\% confidence region of the overall $R_t$ trajectories inferred using the multi-group model (red lines) and the one-group model (light blue lines), when the generation time  for the one-group model is correctly calculated but the contact matrix changes in time. The underlying disease incidence data are generated using a two-group population and the group-specific daily disease incidence data are shown in the top panels of the figure.}
\label{Corrected serial interval, changed Ct}
\end{figure}
In Figure \ref{Corrected serial interval}, we can visually observe that the estimated value of the overall $R_t$ for both the one-group model and the multi-group model align for large values of $t$ when the average number of contacts remains unchanged. However, for a contact matrix that frequently changes over the simulation period, caused by either behavioural changes in mixing patterns or changes to public health measures, the two inferred overall $R_t$ profiles no longer match when the levels of the contact matrix fluctuate according to the entries of Table \ref{tab:changing-ct} (Figure \ref{Corrected serial interval, changed Ct}).
\begin{table}[H]
    \centering
\renewcommand{\arraystretch}{1.3}
\begin{tabular}{|c|c|}
\hline 
\textbf{Time of change}&  \textbf{Change in Contacts ($\%$ Original)}\\ \hline 
         20&  $75\%$\\ \hline 
 40& $50\%$\\\hline
         100&  $80\%$\\ \hline
    \end{tabular}
    \vspace{3mm}
    \caption{Changes in contacts used in our simulation scenarios.}
    \label{tab:changing-ct}
\end{table}

Therefore, in order to accurately estimate the overall $R_t$ using the standard one-group renewal equation model in the context of a structured population with dynamic mixing patterns, we need continuous collection of data from which changes in the contact matrix and generation time can be inferred.

\subsection{Application to real-world outbreaks}

\begin{figure}[ht!]
  \centering
  % include first image
  \begin{subfigure}{0.8\textwidth}
  \centering
\includegraphics[width=.9\linewidth]{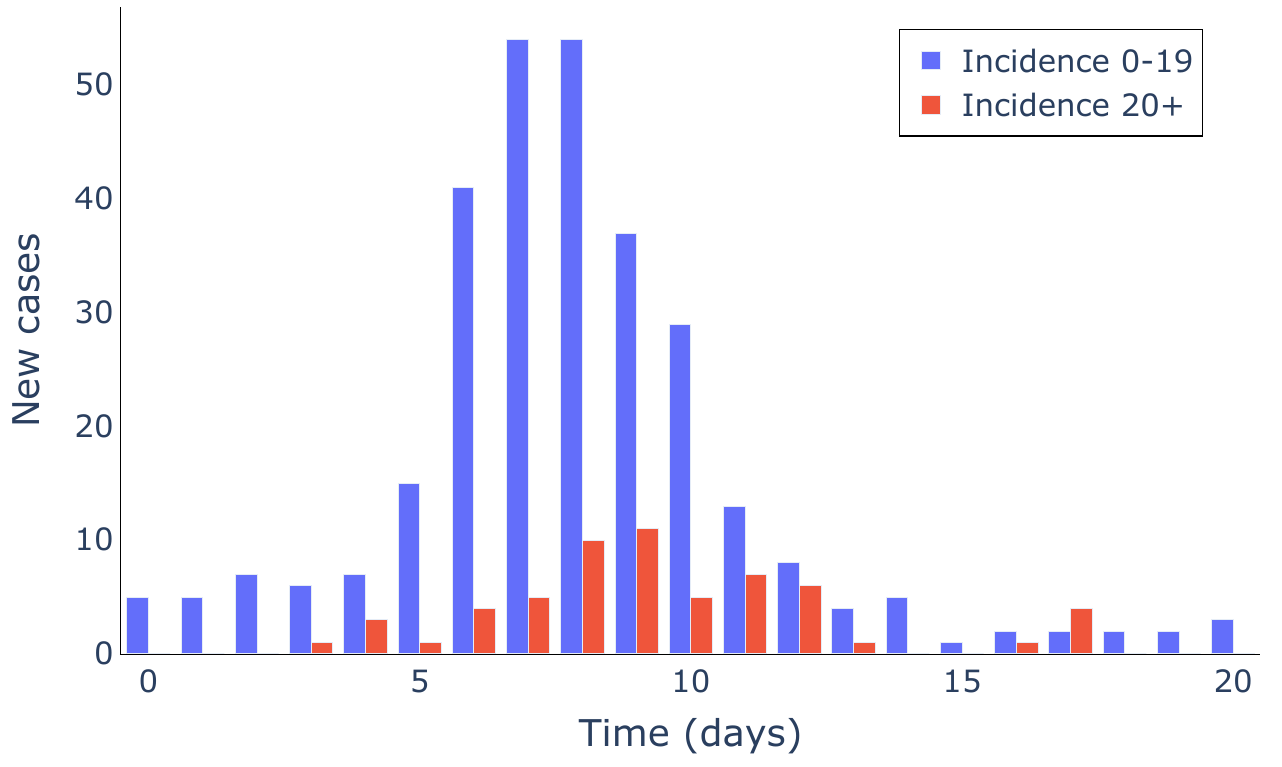}
  \end{subfigure}
  \begin{subfigure}{0.8\textwidth}
\centering
\includegraphics[width=.9\linewidth]{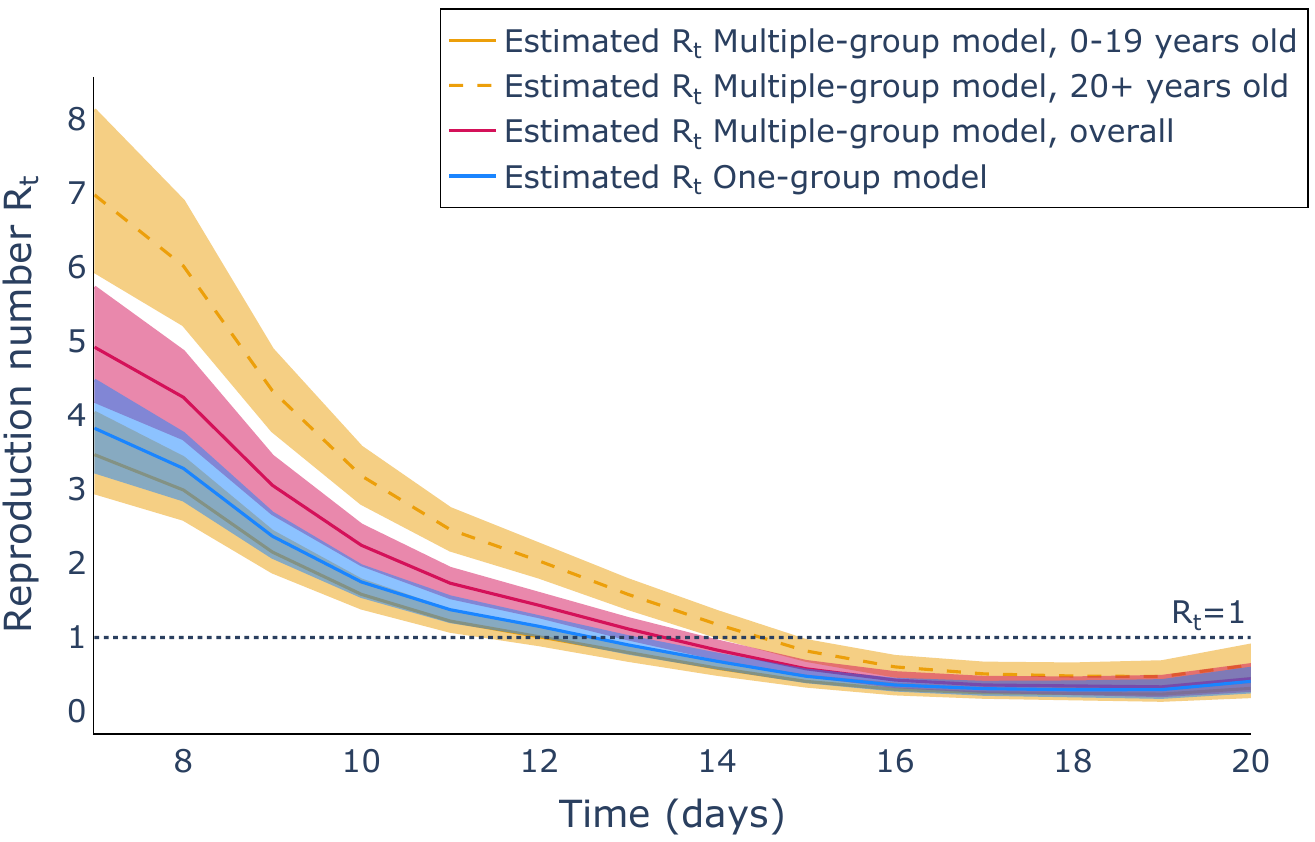}
  \end{subfigure}
\caption{\textbf{$R_t$ estimates obtained using the one-group and multi-group models from disease incidence data from the 2009 Japan A/H1N1 epidemic.} (Top panel) Disease incidence data for the A/H1N1 Japan outbreak in $2009$. The numbers of recorded cases are split across two population groups: 0-19 (blue bars) and $20+$ years old (red bars). (Bottom panel) Inferred mean trajectories and $95\%$ confidence regions of the group-specific reproduction numbers (yellow lines) and overall $R_t$ inferred from the multi-group renewal equation model (red lines) and the one-group model (blue lines). The epidemic dataset consists of data for $20$ days, starting from $9^\text{th}$ May 2009.}
\label{Influenza comparison}
\end{figure}

We now consider data from Japan from the 2009 A/H1N1 influenza epidemic \cite{Nishiura2009b} and investigate how using a multi-group renewal model instead of the standard one-group model changes the estimated $R_t$ trajectory. The disease incidence data are categorised by age, with individuals split into young (0-19 years old) and adults ($20+$ years old), as shown in the top panel of Figure \ref{Influenza comparison}. The recorded number of cases in the young generally exceeds those observed in adults.

In the bottom panel of Figure \ref{Influenza comparison}, the inferred $R_t$ trajectories obtained from the multi-group and one-group renewal equation models are shown. For both age groups considered, we assumed identical gamma-distributed generation times with mean $1.9$ days and standard deviation $0.9$ days \cite{Nishiura2009b}. For the transpose of the effective contact matrix $C_t$ (as stipulated by eq. \eqref{eq:group-specific-incidence} in section \ref{multiple-group versus one-group Models}), we use the contact matrix for Japan \cite{POLYMOD} and $2010$ census data \cite{AH1N1_data} to appropriately weight the contacts in the prescribed age groups, giving
$$ C = \begin{pmatrix}
6.96 & 6.72 \\
1.29 & 9.88
\end{pmatrix}.
$$
\noindent We also assumed that there is no systematic variation in susceptibility or infectiousness between the age groups, aside from transmission varying based on the number of contacts within and between groups, such that the transposed effective contact matrix is exactly the one computed above. The mean and standard deviation of the gamma-prior for $R_t$ are both equal to $5$ days.

We find that the inferred group-specific $R_t$ is generally higher for the $20+$ year-olds ($R_t^{(20+)}$) than for the 0-19 year-olds ($R_t^{(0-19)}$), despite the larger numbers of cases in the young. A possible explanation for this is that there were more infections in the young at the start of the time series in the top panel of Figure \ref{Influenza comparison}, and so -- despite the larger group-specific $R_t^{(20+)}$ in adults -- more infections were seen in children. This can also explain why the overall $R_t$ estimated using the one-group population model more strongly resembles the inferred group-specific reproduction number of the 0-19 year-olds, as the vast majority of recorded infections originate in this particular age group. Comparing the overall $R_t$ trajectories plotted in Figure \ref{Influenza comparison}, we note that using the one-group model, the inferred $R_t$ declines below one slightly earlier than under the multi-group model.

\section{Discussion}
The time-dependent reproduction number, $R_t$, is increasingly inferred and tracked during infectious disease outbreaks \cite{white2021statistical, vegvari2022commentary}. During the COVID-19 pandemic, for example, the value of $R_t$ was estimated in countries and regions worldwide and used to guide policy decisions \cite{Abbott2020}. Estimates of $R_t$ are often obtained by assuming that new cases are generated according to a renewal equation. The simplest renewal equation models involve an assumption that the generation time distribution, which governs the (relative) infectiousness profile during each individuals' infectious period, is the same for all infected individuals. Here, we have developed a Bayesian framework for $R_t$ inference that accounts for the fact that populations are often structured into groups with different characteristics. Specifically, we have demonstrated how $R_t$ can be estimated when different groups within the population are associated with different generation time distributions and accounting for contact structure in the population. We have also shown that, in some scenarios, it is possible to infer $R_t$ accurately under a one-group renewal equation framework, so long as the generation time distribution is chosen carefully. This aggregate generation time distribution is effectively a weighted average of the generation times in each of the population groups. However, a challenge besetting accurate $R_t$ inference in practice is the possibility that the contact matrix varies over time.

Using epidemiological data from Japan from the 2009 A/H1N1 epidemic \cite{Nishiura2009b}, we showed that $R_t$ estimates under the one-group renewal model can be different in practice from those obtained from a model that accounts for age-structure. The practical applicability of the multi-group renewal approach though has limitations -- primarily, its reliance on far richer data compared to the standard one-group model. To use the multi-group $R_t$ inference framework presented here, reliable estimates of the contact matrix are required, ideally including any variation in its structure over time. The multi-group model also requires group-specific case incidence data and, more restrictively, group-specific generation time distributions (or serial interval distributions) to be known. Such data could be determined if fine-scale and dynamic contact-tracing data were accessible; proximity-based mobile phone applications as used in some countries during the COVID-19 pandemic could, in principle, be used to infer such quantities, although data privacy issues complicate this in many locations \cite{parker2020ethics}.

Our extended renewal framework also does not account for imported infections \cite{Thompson2019,Creswell2022}, or differences in reporting between population groups or over time \cite{dalziel2018unreported}. Temporal changes in reporting could occur for a number of reasons, including different symptom development rates across demographies (e.g.\ SARS-CoV-2 infection generally leads to milder symptoms in children, compared to in older adults), the availability of resources (e.g.\ the reporting system progressively getting overwhelmed, especially in remote areas where testing is limited), or behavioural changes during the epidemic (e.g.\ due to public awareness of the outbreak varying, or perceived penalties such as the requirement to isolate associated with reporting disease). It is possible that use of a multi-group modelling framework could amplify these biases.

Renewal equation models will continue to underlie quantitative epidemiological analyses that are used to guide public health policy. They facilitate inference of an explainable metric, $R_t$, that can be used to guide policy decisions. Renewal equation models appear relatively simple, yet this simplicity masks a range of assumptions about the populations they are used to describe. Here, we have probed the assumption of homogeneity of the population within many of these models and showed that models that relax this simple assumption can produce different $R_t$ estimates. The collection of finer-scale epidemic data in future could be used to refine $R_t$ estimates, leading to more informed decision-making during future infectious disease outbreaks.

\section{Acknowledgements}

IB and RNT would like to thank members of the Infectious Disease Modelling group in the Mathematical Institute at the University of Oxford for useful discussions about this work.

\bibliography{References.bib} % local copy
\bibliographystyle{ieeetr}
\newpage

\beginsupplement

\makeatletter
\renewcommand \thesection{S\@arabic\c@section}
\renewcommand\thetable{S\@arabic\c@table}
\renewcommand \thefigure{S\@arabic\c@figure}
\makeatother

\appendix
\setcounter{section}{0}

\section{Appendix}

\begin{figure}[ht!]
  \centering
  % include first image
  \includegraphics[width=.78\linewidth]{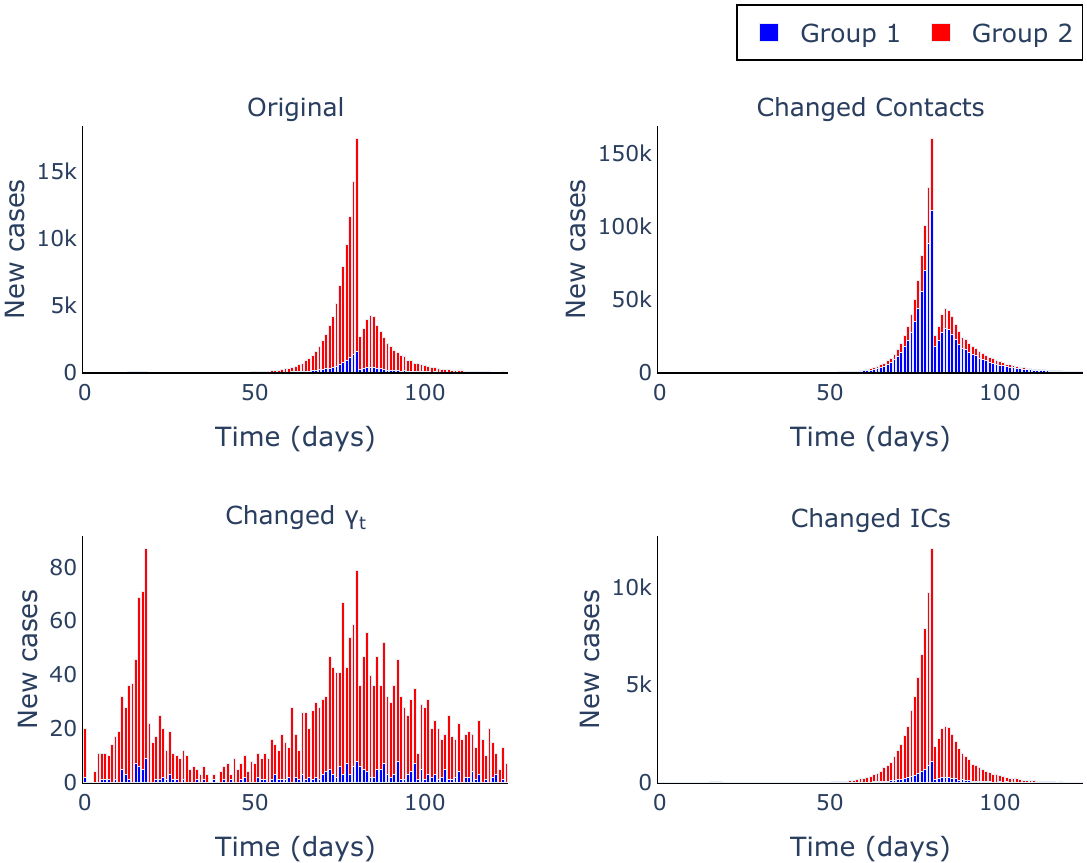}
  \vspace{0.05mm}
\caption{\textit{Disease incidence data generated using a two-group model. These data are used to infer the trajectory of the overall $R_t$ in Figure \ref{Same serial interval} for both the one-group and the multi-group renewal models for a range of changed model parameters as described in Table \ref{tab:comparison-conditions}. The same generation time distribution is used for both population groups. The group-specific mean and standard deviation for the generation time distributions as described in Table \ref{tab:generation-time-comparison}.}}
\label{Same SI Incidence}
\end{figure}
\begin{figure}[ht!]
  \centering
  % include first image
  \includegraphics[width=.78\linewidth]{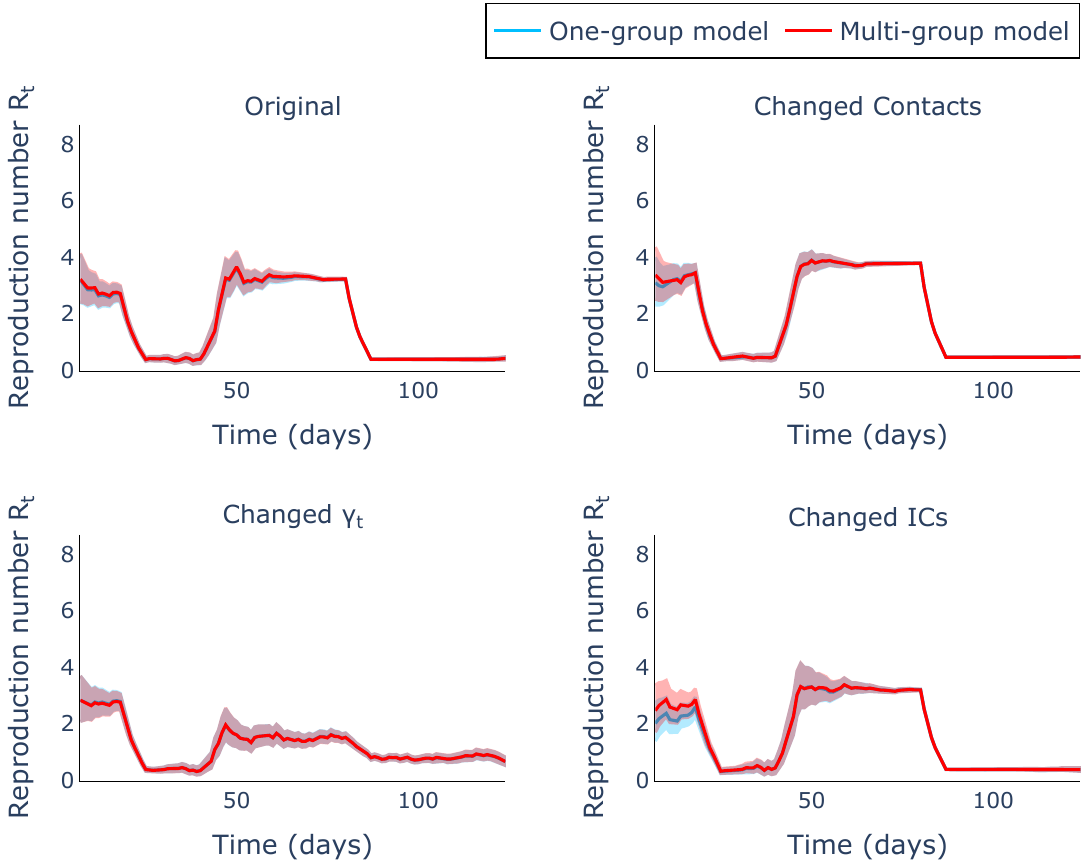}
  \vspace{0.05mm}
\caption{\textit{Comparison of the mean and 95\% confidence regions of the inferred overall $R_t$ trajectories estimated using the multi-group model (red lines) and the one-group model (light blue lines) for a range of changed model parameters as described in Table \ref{tab:comparison-conditions}. The underlying disease incidence data are generated using the two-group model. The same generation time distribution is assumed for both population groups. The group-specific mean and standard deviation for the generation time distributions as described in Table \ref{tab:generation-time-comparison}.}}
\label{Same serial interval}
\end{figure}
\begin{figure}[ht!]
  \centering
  % include first image
  \includegraphics[width=.78\linewidth]{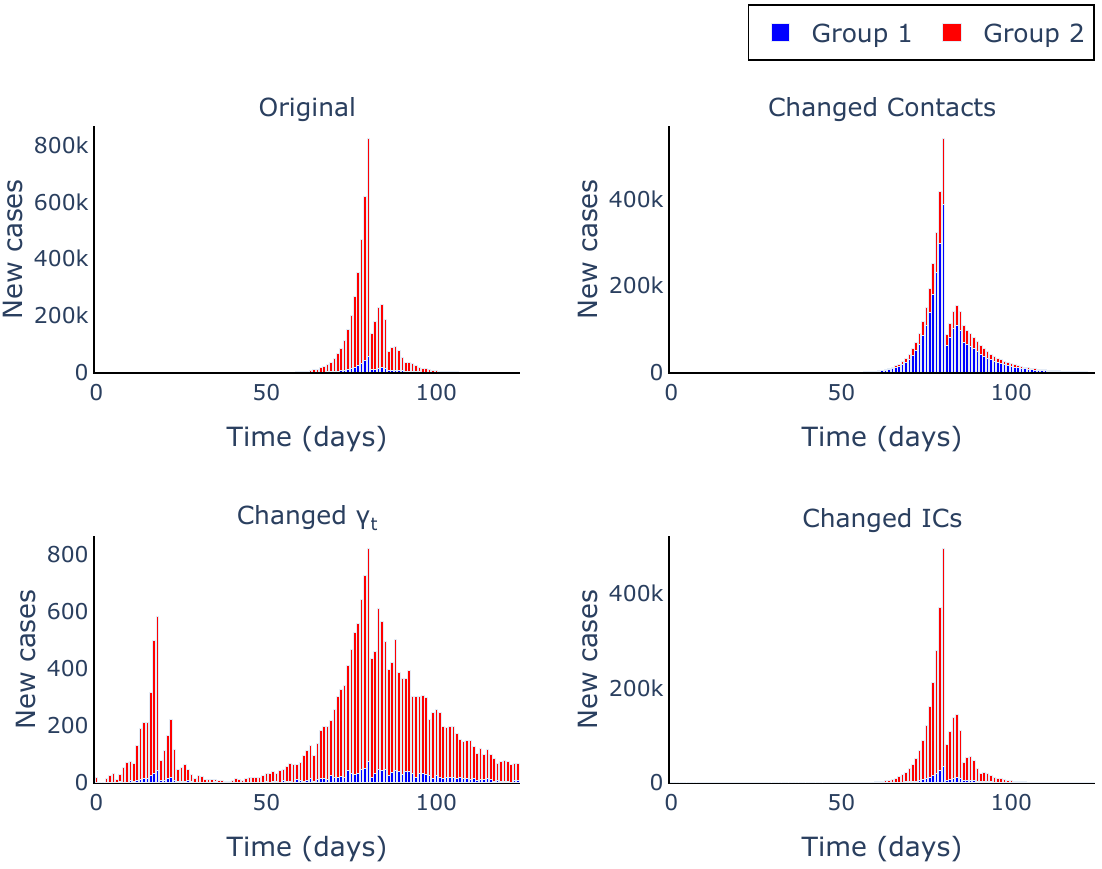}
  \vspace{0.05mm}
\caption{\textit{Disease incidence data generated using a two-group model. These data are used to infer the trajectory of the overall $R_t$ in Figure \ref{Same cropped serial interval} for both the one-group and the multi-group renewal models for a range of changed model parameters as described in Table \ref{tab:comparison-conditions}. The same baseline generation time distribution is used for each of the two population groups, but in the first group the generation time distribution is cut off to reflect interventions applying later in infection. The group-specific mean and standard deviation for the generation time distributions as described in Table \ref{tab:generation-time-comparison}.}}
\label{Same cropped SI Incidence}
\end{figure}
\begin{figure}[ht!]
  \centering
  % include first image
  \includegraphics[width=.78\linewidth]{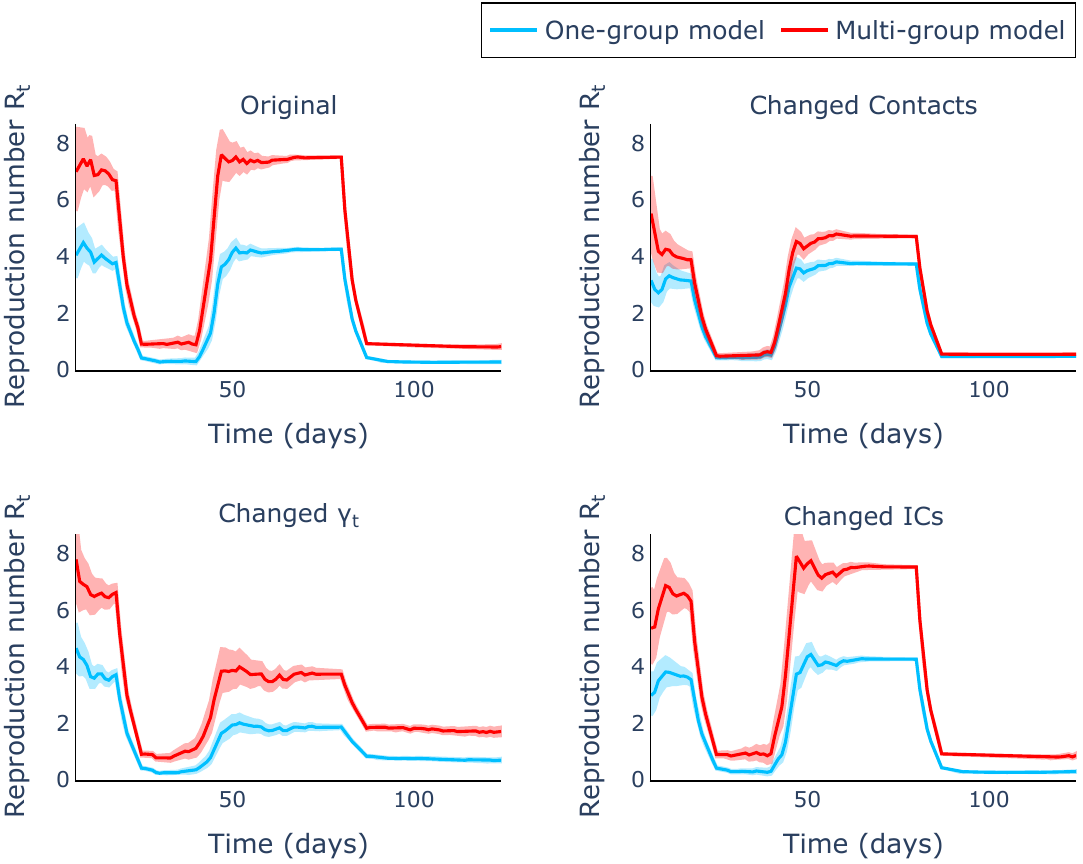}
  \vspace{0.05mm}
\caption{\textit{Comparison of the mean and 95\% confidence regions of the overall $R_t$ trajectories inferred using the multi-group model (red lines) and the one-group model (light blue lines) for a range of changed model parameters as described in Table \ref{tab:comparison-conditions}. The underlying disease incidence data are generated using the two-group model. The same baseline generation time distribution is used for each of the two population groups, but in the first group the generation time distribution is cut off to reflect interventions applying later in infection.}}
\label{Same cropped serial interval}
\end{figure}
\begin{figure}[ht!]
  \centering
  % include first image
  \includegraphics[width=.78\linewidth]{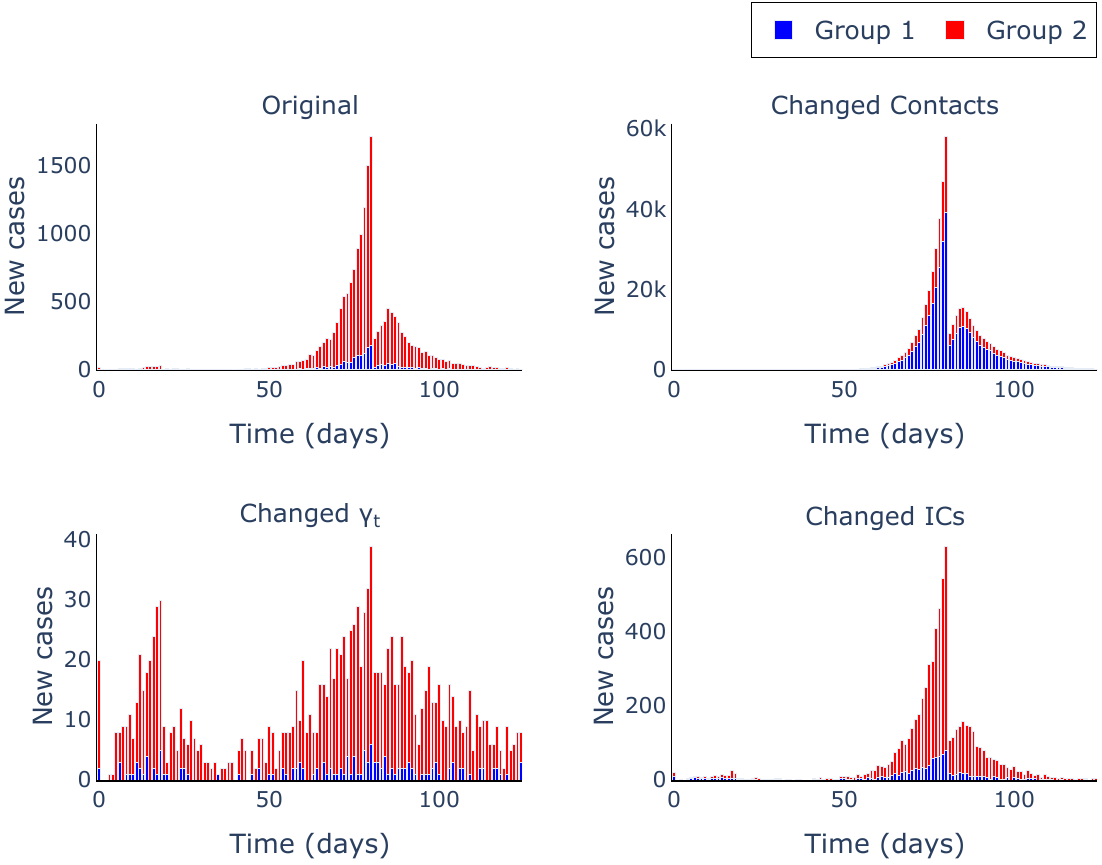}
  \vspace{0.05mm}
\caption{\textit{Disease incidence data generated using a two-group model. These data are used to infer the trajectory of the overall $R_t$ in Figure \ref{Different serial interval} for both the one-group and the multi-group renewal models for a range of changed model parameters as described in Table \ref{tab:comparison-conditions}. Different generation time distributions are used for each population group. The group-specific mean and standard deviation for the generation time distributions as described in Table \ref{tab:generation-time-comparison}.}}
\label{Different SI Incidence}
\end{figure}
\begin{figure}[ht!]
  \centering
  % include first image
  \includegraphics[width=.78\linewidth]{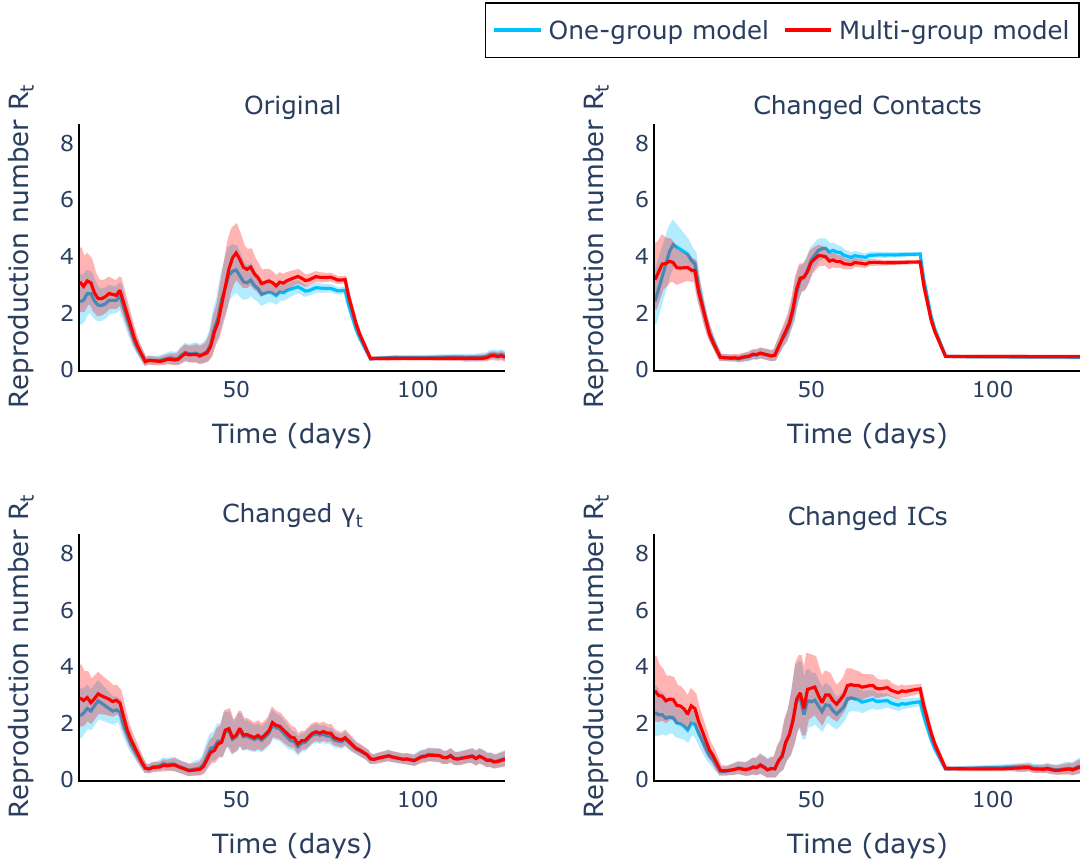}
  \vspace{0.05mm}
\caption{\textit{Comparison of the mean and 95\% confidence regions of the overall $R_t$ trajectories inferred using the multi-group model (red lines) and the one-group model (light blue lines) for a range of changed model parameters as described in Table \ref{tab:comparison-conditions}. The underlying disease incidence data are generated using a two-group model. Different generation time distributions are used for each population group. The group-specific mean and standard deviation for the generation time distributions as described in Table \ref{tab:generation-time-comparison}.}}
\label{Different serial interval}
\end{figure}
\begin{figure}[h!]
  \centering
  % include first image
  \includegraphics[width=.78\linewidth]{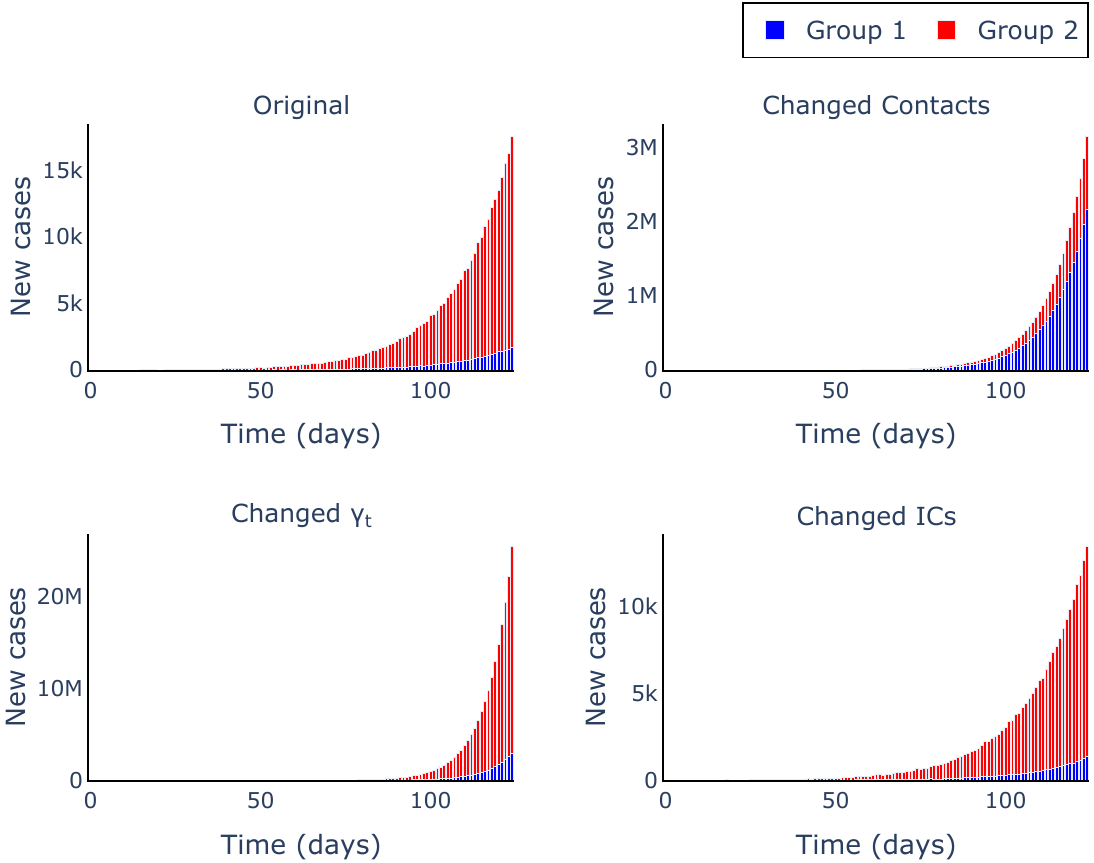}
  \vspace{0.05mm}
\caption{\textit{The underlying case data generated using a two-group population and used to infer the trajectory of the overall $R_t$ in Figure \ref{Corrected serial interval} for both the one-group and the multi-group renewal models, for a range of changed model parameters as described in Table \ref{tab:comparison-conditions}. Different generation time distributions are used for each population group, but the generation time for the one-group model is correctly calculated. The group-specific mean and standard deviation for the generation time distributions as described in Table \ref{tab:generation-time-comparison}.}}
\end{figure}
\begin{figure}[ht]
  \centering
  % include first image
  \includegraphics[width=.7\linewidth]{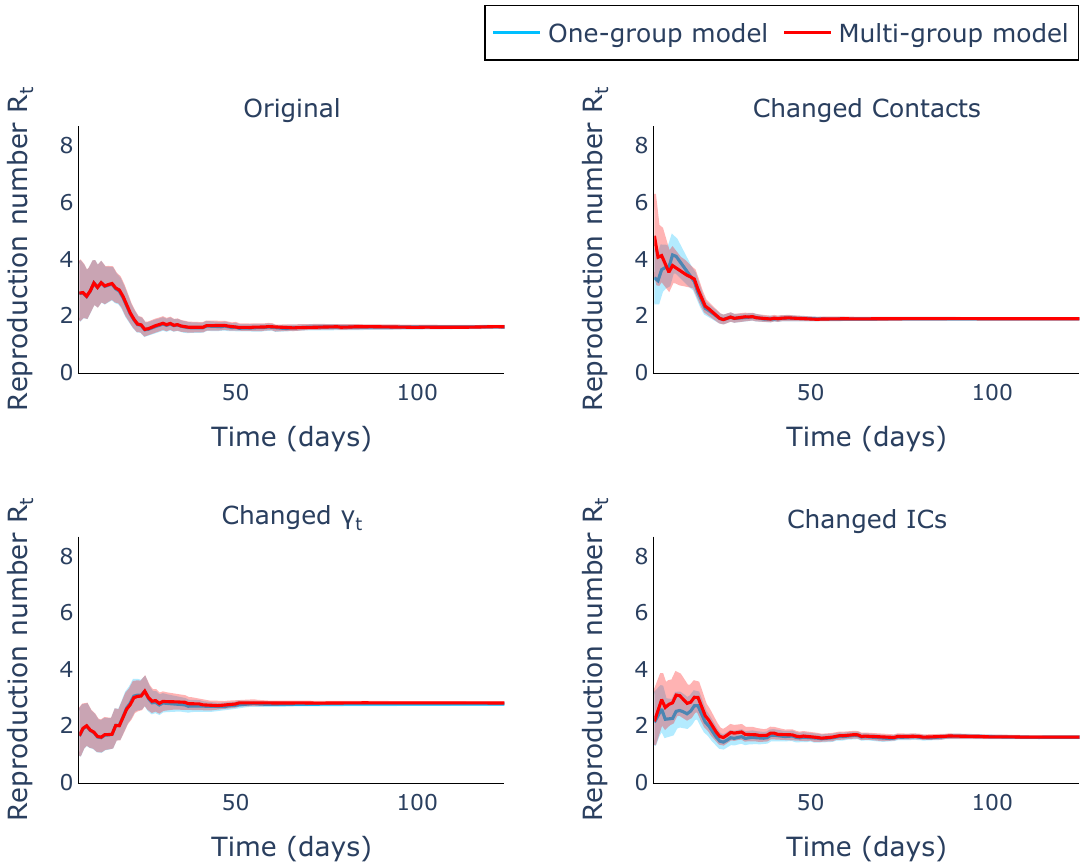}
  \vspace{0.05mm}
\caption{\textit{Comparison of the mean and 95\% confidence region of the overall reproduction number trajectories inferred using the multiple population group model (red lines) versus one population group model (light blue lines) for a range model parameters as described in Table \ref{tab:comparison-conditions}. The underlying disease incidence data are generated using a two-group model. Different generation time distributions are used for each population group, but the generation time for the one-group model is correctly calculated. The group-specific mean and standard deviation for the generation time distributions as described in Table \ref{tab:generation-time-comparison}.}}
\end{figure}
\begin{figure}[ht!]
  \centering
  % include first image
  \includegraphics[width=.78\linewidth]{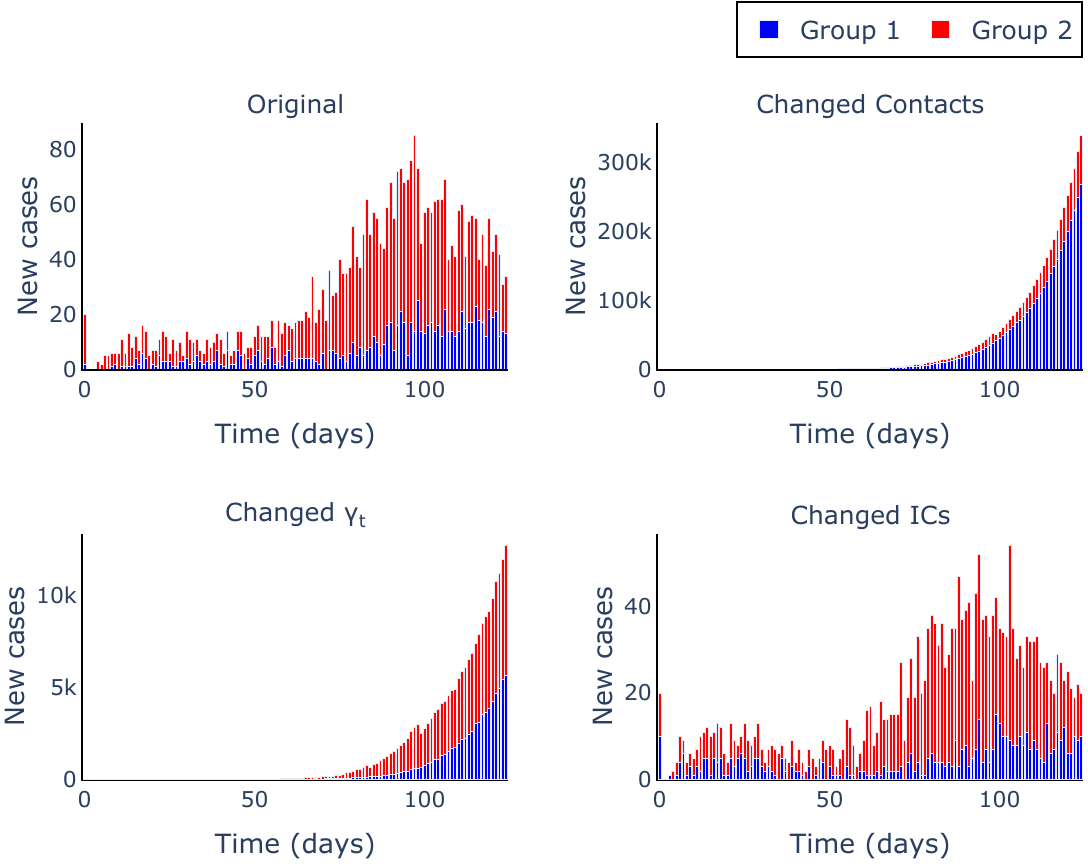}
  \vspace{0.05mm}
\caption{\textit{Disease incidence data generated using a two-group model and used to infer the overall $R_t$ trajectory in Figure \ref{Corrected serial interval, changed Ct} under both the one-group and the multi-group models, for a range model parameters as described in Table \ref{tab:comparison-conditions}. Different generation time distributions are used for each population group, but the generation time for the one-group model is correctly calculated and the contact matrix varies temporally. The group-specific mean and standard deviation for the generation time distributions as described in Table \ref{tab:generation-time-comparison}.}}
\end{figure}
\begin{figure}[ht!]
  \centering
  % include first image
  \includegraphics[width=.7\linewidth]{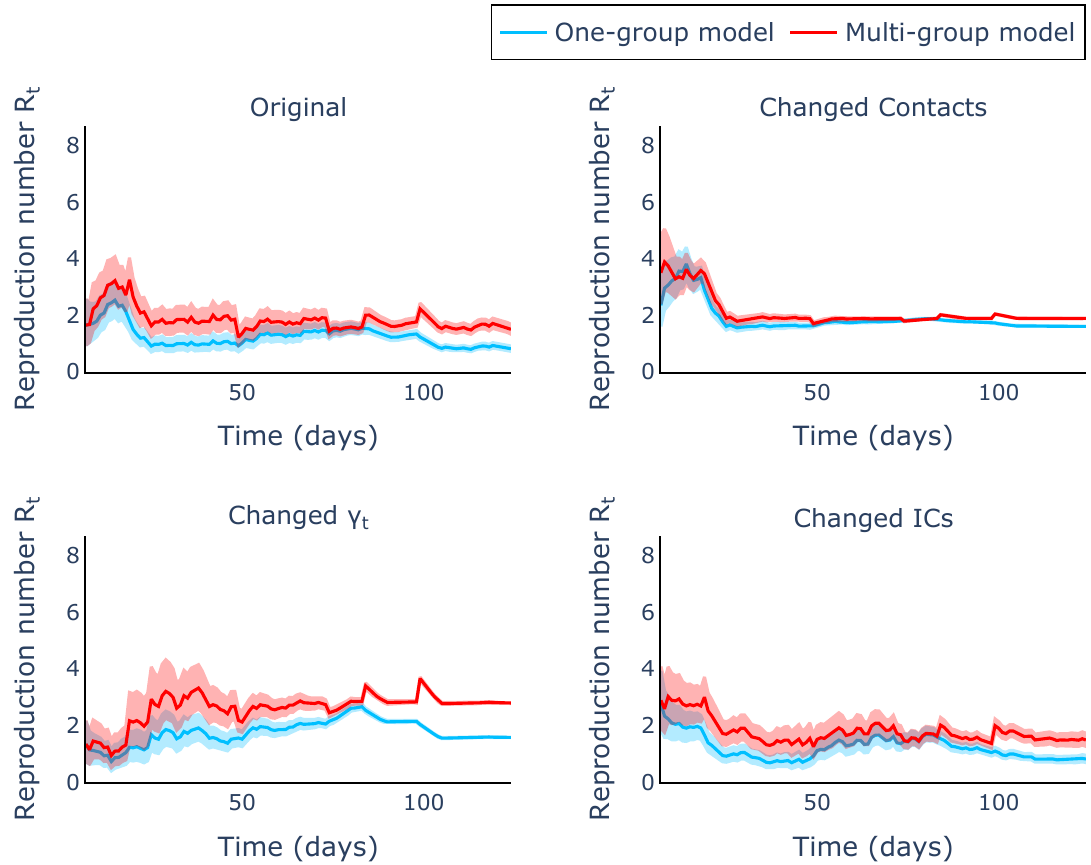}
  \vspace{0.05mm}
\caption{\textit{Comparison of the mean and 95\% confidence region of the overall reproduction number trajectories inferred using the multiple population group model (red lines) versus one population group model (light blue lines) for a range model parameters as described in Table \ref{tab:comparison-conditions}. The underlying disease incidence data are generated using a two-group model. Different generation time distributions are used for each population group, but the generation time for the one-group model is correctly calculated and the contact matrix varies temporally. The group-specific mean and standard deviation for the generation time distributions as described in Table \ref{tab:generation-time-comparison}.}}
\end{figure}

\end{document}